\documentclass[12pt]{iopart}
\usepackage{graphicx}
\usepackage{iopams}
\usepackage[section]{placeins}   
\usepackage{hyperref}  
\usepackage{subfigure}
\hypersetup{colorlinks,urlcolor=blue,linkcolor=blue,citecolor=blue,filecolor=blue,urlcolor=blue}
\usepackage{ifthen}
\usepackage{cite}

\begin{document}

\markboth{V N A Lula-Rocha, S Floquet, M A S Trindade, M C B Fernades and J D M Vianna}
{Negativity of the Wigner function and thermal effects of Bell-Cat states}

\title{Negativity of the Wigner function and thermal effects of Bell-Cat states}  

\author{Vin\'icius N. A. Lula-Rocha$^{1}$, Sergio Floquet$^{2}$, Marco A. S. Trindade$^{3}$, Marco C. B. Fernandes$^{4}$ and Jos\'e D. M. Vianna$^{1,4}$}
\address{$^{1}$ Instituto de F\'isica, Universidade Federal da Bahia, Salvador-BA, 40210-340, Brazil}
\address{$^{2}$ Colegiado de Engenharia Civil, Universidade Federal do Vale do S\~ao Francisco,  Juazeiro-BA, 48902-300, Brazil}
\address{$^{3}$ Colegiado de F\'isica, Departamento de Ci\^encias Exatas e da Terra, Universidade do Estado da Bahia, Salvador-BA, 41150-000, Brazil }
\address{$^{4}$ Instituto de F\'isica, Universidade de Bras\'ilia, Bras\'ilia-DF, 70919-970, Brazil}
\ead{viniciusnonato@gmail.com}

\date{\today}

\begin{abstract}
We applied the Thermofield Dynamics formalism to analyze how the non-classical properties of the Bell-Cat states are influenced by a gradual change of temperature values, in a thermal equilibrium system. To this purpose we calculate the thermal Wigner functions for these states, whose negative volume is associated with non-classical properties, and we evaluate how these non-classical features vary with temperature. Our results indicate that these properties are almost absent for temperatures of around $2\mathrm{K}$.

\vspace{0.5cm}
\end{abstract}

\section{Introduction}  

Years ago, in the celebrated Einstein, Podosky and Rosen's paper \cite{EPR}, attention was drawn to the non-local characteristics associated with the formalism of quantum mechanics. 
Since then, tests of the non-locality of quantum mechanics have been proposed for the analysis and the study of spatially-separated states using theories based on the concept of local realism \cite{Bell}. 
In this context one of the most important analysis is the test of the Bell's inequality, whose violation tipfies non-local properties such as entanglement. 
The maximally entangled Bell states have been widely used mostly due to their experimental evidences which have been known since the 1980s \cite{Aspect}. Nowadays they play a fundamental role on quantum computation, quantum information and quantum cryptography \cite{zelliger}.

Yurker and Stoler \cite{yurker1986,yurker1987} showed that a superposition of coherent states can be obtained by means of the evolution of a coherent state called the Schr\"odinger's  Cat-like state, or only Cat state.
In 1992 Sanders \cite{Sanders} generalized the notion of entangled particle state and introduced the entangled coherent states. Such states can be generate by assembling a macroscopic distinguished superposition of coherent states \cite{leibfried} and the vacuum state in a  Mach-Zehnder interferometer (NLMZI). One feature of theses states in contrast with the Bell states is their non-orthogonality between themselves. Thus theoretical efforts were required to understand the importance of non-orthogonal states in the analysis of the entanglement \cite{hirota2001,gilchrist12}. It turned out that states entangled through non-orthogonal states can reach the maximal entanglement whatever the nature of the states to be entangled \cite{Hirota2001Decoherence}.

The set of the four quasi Bell states based on coherent states is named by quasi Bell entangled states or Bell-Cat states, since it can be produced by Cat states in a NLMZI. 

Teleportation based on coherent states \cite{Hirota2001Teleportation,Kim2001,Wang}, quantum computation \cite{Kim2002,Ralph}, error free quantum reading \cite{HirotaQuantumFreeError} and quantum enigma cipher \cite{Hirota2016}  have been investigated by means of Bell-Cat states. However, few studies involving the thermal effects of these states have been done. A recent study \cite{Kato} analyzed the Bell-Cat states in the presence of thermal noise, and in particular its degree of entanglement and the error performance in the minimax discrimination problem \cite{Hirota2016}. However, it did not offer thorought investigation since thermal noise was assumed affecting only one of the two modes of each state.

In this paper we make a theoretical analysis of the non-classicality of the Bell-Cat states subjected to the influence of temperature. In this new analysis we will use the procedure of Thermofield Dynamics \cite{UmezawaLivro}, in particular the thermal Wigner functions for the Bell-Cat states will be considered along with their negativity properties. The Termofield Dynamics formalism was developed by Umezawa and Takahasi \cite{takahashi,Umezawa} and it is a suitable way to introduce temperature in quantum states \cite{trindade,floquet,kitajima,prudencio,prudencio2}. 

We organized the paper as follows: in Section \ref{sec2} we present the Bell-Cat states. In Section \ref{sec3}  we apply the Thermofield Dynamics formalism to construct the thermal Bell-Cat states and their corresponding thermal density operators. In Section \ref{sec4} we obtain the thermal Wigner functions and analyze their behaviour. We then compute negative volumes of the thermal Wigner functions and establish the relation between temperature and non-classical properties. Finally we present our conclusions in Section \ref{sec5}, followed by the acknowledgments and references.

\section{The Bell-Cat states } 
\label{sec2}

A single mode of a stabilized laser can be described by a coherent state  
\begin{eqnarray}
|\alpha \rangle & = &  
\displaystyle  e^{\displaystyle -\frac{|\alpha|^{2}}{2} }  \sum_{n=0}^{+\infty} \frac{\alpha^{n}}{ \sqrt{ n!}  } |n \rangle,
\end{eqnarray}
where $\alpha \in \mathbb{C}$ is associated with the average field amplitude and $|n \rangle$ is the number Fock state. Yurker and Stoler \cite{yurker1986,yurker1987} showed that a superposition of two coherent states can be obtained by the time evolution of a coherent state, in an amplitude-dispersive medium. These superpositions were called Schr\"odinger's  Cat-like states or Cat states for short. In the form of the superposition of two opposed states, they are written as
\begin{eqnarray}
 |\alpha \rangle_{\pm} & = & \bar{N}_{\pm} \left[ |\alpha \rangle \pm   |-\alpha \rangle \right],
\end{eqnarray}
with $\bar{N}_{\pm} =  \displaystyle \left[2\left(  1 \pm e^{-2|\alpha|^{2}} \right)  \right]^{-\frac{1}{2}} $. By combining Cat states with the vacuum state  on a $50/50$ beam-splitter, the output are the entangled Bell-Cat states \cite{hirota2001,gilchrist12} which are represented by
 \begin{eqnarray}
  \vert \Phi_{\pm}\rangle &=& N_{\pm}
  \left[  |\alpha , \alpha\rangle  \pm |-\alpha , -\alpha\rangle
  \right],  \label{bellCatState1}   \\
  \vert \Psi_{\pm}\rangle &=& N_{\pm}
  \left[  |\alpha , -\alpha\rangle \pm |-\alpha ,\alpha\rangle
  \right],  \label{bellCatState2} 
\end{eqnarray}
where $N_{\pm} =  \displaystyle \left[2\left(  1 \pm e^{-4|\alpha|^{2}} \right)  \right]^{-\frac{1}{2}} $. Bell-Cat states comprise a set of non-orthogonal Bell states and have the  property to recover to the regular orthogonal Bell states for larger values of $\alpha$. For exemple, for $|\alpha|=2$ we have $|\langle \alpha |  -\alpha \rangle |^{2} = \displaystyle e^{-4|\alpha|^{2}} = 1.13\times 10^{-7}$ such that $|\alpha\rangle$ and $|-\alpha\rangle$  can be considered approximately orthogonal states. The Bell states are recovered by encoding logical qubits in the coherents states $|0 \rangle_{L} = |-\alpha \rangle$ and $|1 \rangle_{L} = | \alpha \rangle$ for $|\alpha|\geq 2$. In so doing, the Bell-Cat states (\ref{bellCatState1}) and (\ref{bellCatState2}) can be written as
 \begin{eqnarray}
 \vert \Phi_{\pm}\rangle &=& N_{\pm}
 \left[  |1 \rangle_{L} |1\rangle_{L}  \pm |0 \rangle_{L}| 0\rangle_{L}
 \right],    \\
 \vert \Psi_{\pm}\rangle &=& N_{\pm}
 \left[  |1 \rangle_{L} |0\rangle_{L}\pm |0 \rangle_{L}|1\rangle_{L}
 \right], 
 \end{eqnarray}
 which are in the familiar from of the Bell states. 

We can introduce the compact notation
\begin{eqnarray}
\vert  \psi_{k,\pm} \rangle & = & N_{\pm} \left[ \vert \alpha, k \alpha \rangle \pm \vert -\alpha, -k \alpha \rangle \right] \label{eq-bell}
\end{eqnarray}
for the Bell-Cat states with $k=\pm 1$, that will be useful for a unified description  in the next sections.

\section{Thermofield Dynamics for the Bell-Cat states: an approach through Lie algebras \label{sec3}}

The Thermofield Dynamics (TFD) formalism consists in associating the ensemble average of some observable $\mathcal{A}$ to the expectation value over the state $\vert 0(\beta) \rangle$, namely thermal vacuum state with $\beta=1/K_{b}T$, where $T$ is the temperature of the system in thermal equilibrium and $K_{b}$ is the Boltzmann constant, that is
\begin{eqnarray}
\langle \mathcal{A} \rangle 
& = &
\langle 0(\beta)|\mathcal{A}|0(\beta)\rangle
=
 \frac{1}{Z(\beta)}Tr \left( e^{-\beta H} A \right) \label{eq-1},
\end{eqnarray}
where $Z(\beta) = Tr \left( e^{-\beta H} \right)$ is the partition function of the system described by the Hamiltonian $H$. 

If we suppose that the vacuum state can be expanded in the Fock basis, the temperature information would be encoded in the coefficients of the expansion. We then write
\begin{eqnarray}
\displaystyle \vert 0(\beta) \rangle 
&=&
\sum_{n} g_{n}(\beta) \vert n \rangle,
 \label{vaccumState}
\end{eqnarray}
where $\{g_{n}\}$ is a set of temperature dependent complex functions. From  (\ref{eq-1}) and (\ref{vaccumState}) we have
\begin{eqnarray}
 \langle A \rangle 
 & = &
  \frac{1}{Z(\beta)} \sum_{n}  e^{-\beta E_{n}} \langle n \vert A \vert n \rangle 
  \nonumber 
  \\
 & = &
 \sum_{n,m} g_{m}^{*}(\beta)  g_{n}(\beta) \langle m \vert A \vert n \rangle,
\end{eqnarray}
with $\displaystyle g_{m}^{*}(\beta)  g_{n}(\beta) = \frac{1}{Z(\beta)}  e^{-\beta E_{n}} \delta_{nm} $, which turns out to be an absurd since this relation cannot be satisfied by $c$-numbers. One possibility to get over this issue is to describe the thermal vacuum state in a doubled Hilbert space $\mathbb{H}_{T}=\mathbb{H} \otimes \widetilde{\mathbb{H}}$ as $\displaystyle \vert 0(\beta) \rangle = \sum_{n} g_{n}(\beta) \vert n  \rangle \otimes \vert \widetilde{n}  \rangle $, where $\widetilde{\mathbb{H}}$ is a replic of the usual Hilbert space $\mathbb{H}$ of the system. The space $\mathbb{H}_{T}$ is also known as thermal Hilbert space \cite{UmezawaLivro}. Hence, the thermal  vacuum state can be written as
\begin{eqnarray}
 \vert 0(\beta) \rangle 
 & = & 
 \frac{1}{\sqrt{Z(\beta)}} 
 \sum_{n} 
  e^{-\frac{\beta E_{n}}{2}} \vert n, \widetilde{n}  \rangle .
\end{eqnarray}

We can introduce an unitary transformation that associates the doubled vacuum state $ \vert 0, \widetilde{0}  \rangle  $ with the thermal vacuum state $\vert 0(\beta) \rangle $, namely the Bougoliubov transformation 
\begin{eqnarray}
 \vert 0(\beta) \rangle  =  U(\beta)  \vert 0, \widetilde{0}  \rangle  ,
\end{eqnarray}
that leads to the definition  $A(\beta) = U(\beta) A U^{\dagger}(\beta)$ of the thermal operators which preserve the structure of the Hilbert space.

We are interested in Bell-Cat states as discribed in the TFD formalism applied to the two modes bosonic harmonic oscillator, described by the Hamiltonian
\begin{eqnarray}
 H & = &  \hslash\omega_{1}a_{1}^{\dagger}a_{1} +  \hslash\omega_{2}a_{2}^{\dagger}a_{2},
\end{eqnarray}
where  $\omega_{i}$ is the angular frequency, $a_{i}$ and $a_{i}^{\dagger}$ the annihilation and creation operators, respectively. They satisfy the usual commutation relations $\left[ a_{i},a_{j}^{\dagger}\right] =1$ and $\left[ a_{i},a_{j}\right] = \left[ a_{i}^{\dagger},a_{j}^{\dagger}\right] = 0$, for the modes $i,j\in\lbrace 1,2 \rbrace$. The Hilbert space $\mathcal{H}$ is now expanded by the two modes number states $|n_{1},n_{2}\rangle$ and following the TFD description we double the Hilbert space. The replic $\mathcal{ \widetilde{H}}$ of the original Hilbert space is expanded by the states $|\tilde{n}_{1},\tilde{n}_{2}\rangle$.

We have annihilation and creation operators in the tilde Hilbert space satisfying analogous commutation relations 
 $\left[ \widetilde{ a_{i}}, \widetilde{ a_{j}}^{\dagger}\right] =1$, $\left[ \widetilde{ a_{i}}, \widetilde{ a_{j}}\right] =\left[ \widetilde{ a_{i}}^{\dagger}, \widetilde{ a_{j}}^{\dagger}\right] =0$.  In addiction all tilde and non-tilde operators commute among themselves. The thermal Hilbert space  for the two modes bosonic oscillator is given by $\mathcal{H}_{T}=\mathcal{H}\otimes\tilde{\mathcal{H}}$ and the thermal vacuum state is 
\begin{eqnarray}
 \vert 0(\beta) \rangle 
 & = & 
 \frac{1}{\sqrt{Z(\beta)}}
 \sum_{n_{1},n_{2}=0}^{\infty}
 e^{-\frac{\displaystyle \beta \hslash\left( \omega_{1}n_{1} + \omega_{2}n_{2}\right) }{\displaystyle 2}} \vert n_{1},n_{2}, \widetilde{n_{1}},\widetilde{n_{2}}  \rangle \nonumber \\
 & = & U(\beta) \vert 0,0,\widetilde{0}, \widetilde{0} \rangle,
\end{eqnarray}
with $Z(\beta) = \displaystyle \frac{1}{1-e^{-\beta\hslash \omega_{1}}} \frac{1}{1-e^{-\beta\hslash \omega_{2}}}$. The Bougoliubov transformation \cite{UmezawaLivro} is then
\begin{eqnarray}
U(\beta) 
& = & 
e^{\displaystyle \theta_{1}(\beta) \left(  \widetilde{a_{1}}^{\dagger}a_{1}^{\dagger}  - \widetilde{a_{1}}a_{1} \right) +  \theta_{2}(\beta) \left(\widetilde{a_{2}}^{\dagger}a_{2}^{\dagger} - \widetilde{a_{2}}a_{2}  \right) },\label{eq-a-ter}
\end{eqnarray}
where $\theta_{i}$, $i\in \{1,2\}$, are $\beta$-dependent parameters defined by the relations
\begin{eqnarray}
\cosh \theta_{i}(\beta) 
& = &
 \frac{\displaystyle 1}{\displaystyle \sqrt{1- e^{-\beta \hslash\omega_{i}} }} \ \equiv \ u_{i}(\beta),
 \label{u_i_beta} 
 \\
\sinh \theta_{i}(\beta) 
& = &
 \frac{\displaystyle e^{-\beta\hslash\omega_{i}/2}}{\displaystyle \sqrt{1- e^{-\beta \hslash\omega_{i}} }} \ \equiv \ v_{i}(\beta). \label{v_i_beta}
 \
\end{eqnarray}

Using (\ref{eq-a-ter}) the  thermal creation and annihilation operators can be written in terms of the non-thermal ones, that is \cite{UmezawaLivro}
\begin{eqnarray}
 a_{i}(\beta) 
 & = &
 u_{i}(\beta) a_{i} - v_{i}(\beta) \widetilde{a_{i}}^{\dagger},
 \label{eq-term-1}
 \\
 a_{i}^{\dagger}(\beta) 
 & = &
 u_{i}(\beta) a_{i}^{\dagger} - v_{i}(\beta) \widetilde{a_{i}},
 \label{eq-term-2} 
 \\
 \widetilde{ a_{i}}(\beta) 
 & = &  
 u_{i}(\beta) \widetilde{a_{i}} - v_{i}(\beta) {a_{i}}^{\dagger}, 
 \label{eq-term-3}
 \\
\widetilde{ a_{i}}^{\dagger}(\beta) 
& = & 
u_{i}(\beta) \widetilde{a_{i}}^{\dagger} - v_{i}(\beta) a_{i}.  
\label{eq-term-4}
\end{eqnarray}
From (\ref{eq-term-1}--\ref{eq-term-4}) we can express a general thermal state $\vert \Psi (\beta) \rangle $ in terms of the thermal vacuum state to make explicit the thermal effects, that is,
\begin{eqnarray}
 \vert \Psi (\beta) \rangle & = & f(a_{i},a_{i}^{\dagger};\beta) \vert 0 (\beta) \rangle \label{generalState},
\end{eqnarray}
where $f$ is a function of the non-tilde operators and of the temperature dependent functions $u_{i}(\beta)$ and $v_{i}(\beta)$, defined in (\ref{u_i_beta}) and (\ref{v_i_beta}).

Now we are able to associate the general thermal state (\ref{generalState}) with the thermal density operator $\rho_{|\Psi(\beta)\rangle} $ using the relation between ensemble average and the expectation value of some operator $O$, i.e.,
\begin{eqnarray} 
\langle \Psi(\beta)|O|\Psi(\beta)\rangle
 &= & Tr(\rho_{|\Psi(\beta) \rangle }O) 
 \nonumber 
 \\
 & = &\langle 0(\beta)|f^{\dagger}(a,a^{\dagger};\beta)Of(a,a^{\dagger};\beta)|0(\beta)\rangle 
 \nonumber 
 \\
&=& Tr(\rho_{\beta}f^{\dagger}(a,a^{\dagger};\beta)Of(a,a^{\dagger};\beta))
 \nonumber 
  \\
&=& 
Tr(f(a,a^{\dagger};\beta)\rho_{\beta} f^{\dagger}(a,a^{\dagger};\beta)O), 
	\label{thermalDensityMatrixDeduction}
\end{eqnarray}
where in the last step of (\ref{thermalDensityMatrixDeduction}) we use the cyclic property of the trace. We can now compare the first and the last lines of the above equation and note that
\begin{equation} \displaystyle
  \rho_{|\Psi(\beta)\rangle}=f(a,a^{\dagger};\beta)\rho_{\beta} f^{\dagger}(a,a^{\dagger};\beta),  \label{eq-op-den}
\end{equation}
where $\rho_{\beta}$ is the density operator of associated with the vacuum state and is given by
\begin{eqnarray}
\rho_{\beta}
&=&
\frac{e^{-\beta H}}{Z(\beta)}.
 \label{thermalDensityVacuum}
\end{eqnarray}

An algebraic construction  of the thermal Bell-cat states can be done by considering the thermal Lie algebra of the Heisenberg-Weyl group $H_{4}^{1}(\beta) \oplus  H_{4}^{2}(\beta)$ \cite{zhang} with the generators $\left\lbrace N_{1}(\beta), a_{1}^{\dagger}(\beta), a_{1}(\beta), I_{1}, N_{2}(\beta), a_{2}^{\dagger}(\beta), a_{2}(\beta), I_{2} \right\rbrace$ and the stability subgroup $\overline{U}^{1}_{\beta}(1) \otimes \overline{U}^{1}_{\beta}(1) \otimes \overline{U}^{2}_{\beta}(1)\otimes \overline{U}^{2}_{\beta}(1) $ whose Lie algebra is spanned by $\left\lbrace N_{1}(\beta), I_{1}, N_{2}(\beta), I_{2} \right\rbrace$. The coset space is
\begin{eqnarray}
 \frac{H_{4}^{1}(\beta) \oplus  H_{4}^{1}(\beta)}{
 	 \overline{U}^{1}_{\beta}(1) \otimes \overline{U}^{1}_{\beta}(1) \otimes \overline{U}^{2}_{\beta}(1)\otimes \overline{U}^{2}_{\beta}(1)
},
\end{eqnarray}
where its representative is the thermal displacement operator
\begin{eqnarray}
 D(\alpha_{1},\alpha_{2} ; \beta) & = & e^{\displaystyle \alpha_{1} a^{\dagger}_{1}(\beta) - \alpha_{1}^{*} a_{1}(\beta) + \alpha_{2} a^{\dagger}_{2}(\beta) - \alpha_{2}^{*} a_{2}(\beta) } \nonumber \\
   & = &   e^{\displaystyle \alpha_{1} a^{\dagger}_{1}(\beta) - \alpha_{1} a_{1}^{*}(\beta)  } e^{\displaystyle  \alpha a^{\dagger}_{2}(\beta) - \alpha^{*} a_{2}(\beta) } \nonumber \\
   & = & D_{1}(\alpha ; \beta) D_{2}(\alpha ; \beta),
\end{eqnarray}
where $\alpha_{1}$ and $\alpha_{2}$ are complex parameters and $ D_{1}(\alpha ; \beta)$ and $ D_{2}(\alpha ; \beta)$ are the thermal displacement operators for the two modes. In particular, if we consider $\alpha_{1}=\alpha_{2}=\alpha$ we obtain $ D(\alpha,\alpha ; \beta)$ and similarly $ D(\alpha,-\alpha ; \beta) $, $ D(-\alpha,\alpha ; \beta) $ and $ D(-\alpha,-\alpha ; \beta) $.

Following this algebraic construction the thermal displacement operator associated with the thermal Bell-cat states are
\begin{eqnarray}
D_{| \Phi_{\pm} \rangle }(\beta) & = & D_{1}(\alpha ; \beta) D_{2}(\alpha ; \beta) \pm D_{1}(-\alpha ; \beta) D_{2}(-\alpha ; \beta), \nonumber \\
D_{| \Psi_{\pm} \rangle }(\beta) & = & D_{1}(\alpha ; \beta) D_{2}(-\alpha ; \beta) \pm D_{1}(-\alpha ; \beta) D_{2}(\alpha ; \beta) ,
\end{eqnarray}
which implies that
\begin{eqnarray}
\hspace{-0.5cm}
 | \Phi_{\pm}(\beta) \rangle & = & N_{\pm} \left[ D_{1}(\alpha ; \beta) D_{2}(\alpha ; \beta) \pm D_{1}(-\alpha ; \beta) D_{2}(-\alpha ; \beta)  \right] |0(\beta),0(\beta) \rangle,
 \nonumber 
  \\
  \hspace{-0.5cm} 
 | \Psi_{\pm}(\beta) \rangle & = & N_{\pm} \left[ D_{1}(\alpha ; \beta) D_{2}(-\alpha ; \beta) \pm D_{1}(-\alpha ; \beta) D_{2}(\alpha ; \beta)  \right] |0(\beta),0(\beta) \rangle .
 \end{eqnarray}

Using the notation of equation (\ref{eq-bell}) and that 
\begin{eqnarray}
\displaystyle D_{i}(\alpha; \beta) |0(\beta)\rangle  \  = \   U(\beta) e^{-\frac{1}{2}|\alpha|^{2}} \sum_{n=0}^{\infty} \frac{\alpha^{n} }{\sqrt{n!} }  \vert n_{i}, \widetilde{0}  \rangle, 
\end{eqnarray}
we perform the Bogoliubov transformation onto the state $|n , \tilde{0} \rangle  \in \mathcal{H}\otimes \widetilde{\mathcal{H}} $, where $|\tilde{0} \rangle $ is the  vacuum state in the tilde Hilbert space $\tilde{\mathcal{H}}$. We double the Bell-Cat states with the vacuum state to ensure the recovering of the original state when  $ T \rightarrow 0$ \cite{trindade,floquet}. The thermalized Bell-Cat states are then
\begin{eqnarray}
\hspace{-1cm}
\vert  \psi_{k,\pm} (\beta) \rangle
 & = & 
N_{\pm}  U(\beta)  
\left[  \vert \alpha, k \alpha, \widetilde{0},\widetilde{0}  \rangle \pm \vert -\alpha, -k \alpha, \widetilde{0},\widetilde{0}  \rangle  \right]  \nonumber   \\
& = &  N_{\pm} e^{-|\alpha|^{2}} \sum_{n,m=0}^{\infty} \frac{\alpha^{n+m} k^{m} }{\sqrt{n! m!} }   \left[
 1 \pm (-1)^{n+m}   \right] 
U(\beta) \vert n, m, \widetilde{0},\widetilde{0}  \rangle   \nonumber  \\
& = & 
N_{\pm} e^{-|\alpha|^{2}} 
\!\!\!\!
\sum_{n,m=0}^{\infty} 
\!\!\!\!
\frac{\alpha^{n+m} k^{m} }{n! m! } 
\!
\left[
 1 \pm (-1)^{n+m}  
\right]
\!\! 
\frac{(a_{1}^{\dagger})^{n} (a_{2}^{\dagger})^{m} }{(u_{1}(\beta))^{n} (u_{2}(\beta))^{m}} \vert 0(\beta)  \rangle.
 \label{eq-22}
\end{eqnarray}
It can be noticed in (\ref{eq-22}) that $|\psi_{k,+}(\beta)\rangle$ are different from zero only in the cases where $n+m$ are even. In another hand $|\psi_{k,-}(\beta)\rangle$ are different from zero for $n+m$ odd.  We can now call $|\psi_{1,+}(\beta)\rangle$ and $|\psi_{-1,+}(\beta)\rangle$ even thermal Bell-Cat states and $|\psi_{1,-}(\beta)\rangle$  and $|\psi_{-1,-}(\beta)\rangle$ odd thermal Bell-Cat states. 

The thermal density operators $\rho_{\vert \psi_{k,\pm}(\beta)\rangle }$ that carry the thermal properties of the  Bell-Cat states can be obtained by using (\ref{thermalDensityMatrixDeduction}) and (\ref{eq-22}) resulting in
\begin{eqnarray}
\hspace{-1cm}
\rho_{\vert \psi_{k,\pm}(\beta)\rangle }
 & = & 
 \frac{
e^{-2|\alpha|^{2}} \left( 1- e^{-\beta \hslash\omega_{1} }\right) \left( 1- e^{-\beta \hslash\omega_{2} }\right)}{ 2\left(  1 \pm e^{-4|\alpha|^{2}} \right)  }
\sum_{\{n,m\}}^{\infty}
 \frac{  \alpha^{n+m}   (\alpha^{*})^{\bar{n} + \bar{m}} k^{m+\bar{m}}  
}{ n! \ \ m! \ \ \bar{n}! \ \ \bar{m}! \ \ n_{1}! \ \ n_{2}!  } \nonumber \\
& &
 \times
 \left[ 1 \pm (-1)^{n+m} \right]
 \left[ 1 \pm (-1)^{\bar{n}+ \bar{m}} \right] 
 \left(
  \sqrt{1 - e^{-\beta \hslash\omega_{1}}}  
 \right)^{n+\bar{n}}  
 \nonumber \\
& &
\times
 \left(
 \sqrt{1 - e^{-\beta \hslash\omega_{2}}}  
 \right)^{m+\bar{m}}
\left( 
 e^{-\beta\hslash \omega_{1}}
\right)^{n_{1}} 
\left( 
 e^{-\beta\hslash \omega_{2}} 
\right)^{n_{2}} 
\sqrt{(n_{1} +n)! \ \ (n_{2} +m)!  }
\nonumber
\\
&&
\times 
\sqrt{(n_{1} +\bar{n})! \ \ (n_{2} +\bar{m})!}
\vert n_{1} + n, n_{2} + m\rangle  \langle n_{1} + \bar{n}, n_{2} + \bar{m} \vert . \label{densityMatrix}
\end{eqnarray}
where we use the symbol $\sum\limits_{\{n,m\}}^{\infty}$ as short notation for  $\sum\limits_{n=0}^{\infty}\sum\limits_{m=0}^{\infty}\sum\limits_{\bar{n}=0}^{\infty}\sum\limits_{\bar{m}=0}^{\infty}\sum\limits_{n_{1}=0}^{\infty}\sum\limits_{n_{2}=0}^{\infty}$.	
From this expression we can observe that the non-zero components of the thermal density operator associated with the Bell-Cat states only remain for both $n+m$ and $\bar{n}+\bar{m}$ even or odd. 

We finish this section by highlighting that the thermal density operators (\ref{densityMatrix}) carry all the information associated with the thermal Bell-Cat states. In another words, they describe theoretically the thermal effects of the physical Bell-Cat states.

\section{Wigner Functions of the thermal Bell-Cat states and their negative volume }
\label{sec4}

A representation of a quantum state in phase space can be made by  evaluating the Wigner Function (WF) \cite{Wigner}, that is a quasi-probability distribution admitting negative values. This negativity of WF is associated with the non-classical properties of quantum states such as non-locality and entanglement. This feature of the WF is useful in order to caracterize theoretically and experimentally states in quantum optics since the negativity of the WF can be measured empirically \cite{bell87,mc15,smi93,ku97}.

The WF is defined as a Fourier-like transformation of the one mode density \mbox{operator $\rho$} 
\begin{eqnarray}
W(q,p)
&=&
\frac{1}{2\pi\hslash}
\int_{-\infty}^{+\infty}
dv
e^{i\frac{vp}{\hslash}}
\langle q -\frac{v}{2}|\rho | q + \frac{v}{2}\rangle.
 \label{wignerFunctionDefinition}
\end{eqnarray}
From this representation, it can be shown that the WF associated with the thermal density operators (\ref{densityMatrix}) is given by the product of the WF of the modes of thermal Bell-Cat states, that is
\begin{eqnarray}
 W_{\alpha,\beta}^{k,\pm}(q_{1},p_{1},q_{2},p_{2})
 &=&
 W_{\alpha,\beta;1}^{k,\pm}(q_{1},p_{1})W_{\alpha,\beta;2}^{k,\pm}(q_{2},p_{2})
 \nonumber
 \\
 &=& 
  \frac{1}{(2\pi \hslash)^{2}} 
  \int_{-\infty}^{+\infty}
   dv_{1} dv_{2}
  \displaystyle 
  \ e^{ i \frac{ v_{1}p_{1}}{\hslash} }  e^{i \frac{v_{2}p_{2}}{\hslash} } 
  \nonumber
  \\
  &&
  \times
  \langle q_{1} - \frac{v_{1}}{2},q_{2} - \frac{v_{2}}{2}| \rho_{|\psi_{k,\pm}(\beta)\rangle} | q_{1} + \frac{v_{1}}{2}, q_{2} + \frac{v_{2}}{2} \rangle,
  \label{wignerFunction}
\end{eqnarray}
where $W_{\alpha,\beta;i}^{\pm}(q_{i},p_{i})$, $i=1,2$, is the WF relative to the $i$th mode of the thermal Bell-Cat states.

A straightforward calculation of (\ref{wignerFunction}) can be performed by using the representation in terms of Hermite polynomials whose integrals lead to the associated Laguerre polynomials $L^{m}_{n}(x)$ \cite{integrais,butkov}. We then obtain the following expression for the Wigner function associated with the thermal Bell-Cat states:
\begin{eqnarray}
\hspace{-2cm}  
W_{\alpha,\beta}^{k,\pm}(q_{1},p_{1},q_{2},p_{2}) 
& = &
\frac{ \left( 1- e^{-\beta\hslash \omega_{1} }\right) \left( 1- e^{-\beta \hslash\omega_{2} }\right) 
e^{ \displaystyle  -\frac{q_{1}^{2}}{b^{2}_{1}} - \frac{p_{1}^{2} b^{2}_{1} }{\hslash^{2}}  } 
e^{ \displaystyle  -\frac{q_{2}^{2}}{b^{2}_{2}} - \frac{p_{2}^{2} b^{2}_{2} }{\hslash^{2}}  } 
}{2\pi^{2} \hslash^{2} \left(  e^{2|\alpha|^{2}} \pm e^{-2|\alpha|^{2}} \right)  }
\nonumber  
\\
&& 
\times
\sum_{\{n,m\}}^{\infty}
\frac{  \alpha^{n+m}   (\alpha^{*})^{\bar{n} + \bar{m}} k^{m+\bar{m}} 
}{ n! \ \ m! \ \ \bar{n}! \ \ \bar{m}! \ \ n_{1}! \ \ n_{2}!  } 
\left[ 1 \pm (-1)^{n+m} \right] 
\left[ 1 \pm (-1)^{\bar{n}+ \bar{m}} \right]
\nonumber 
\\
\hspace{-0.5cm}
&&
\times
\left(\sqrt{1 - e^{-\beta\hslash \omega_{1}}}  \right)^{n+\bar{n}}  
\left(\sqrt{1 - e^{-\beta \hslash\omega_{2}}}  \right)^{m+\bar{m}}  
\left( e^{-\beta\hslash \omega_{1}} \right)^{n_{1}} 
\nonumber
\\
&&
\times 
\left( e^{-\beta\hslash \omega_{2}} \right)^{n_{2}}
(n_{1} + min(n,\bar{n}))! 
(n_{2} + min(m,\bar{m}))!
(-1)^{n_{1} + n}
\nonumber
\\
&&
\times   
(-1)^{  \displaystyle  n_{2} + m - min(0,n-\bar{n}) - min(0,m-\bar{m}) }
(\sqrt{2})^{|n-\bar{n}|} 
\nonumber
\\
&&
\times  
(\sqrt{2})^{|m-\bar{m}|}
\chi_{1}^{|n-\bar{n}|}  \ \chi_{2}^{|m-\bar{m}|} \ 
L^{|n-\bar{n}|}_{n_{1} + min(n,\bar{n})} \left(\frac{2q_{1}^{2}}{b^{2}_{1}} + \frac{2p_{1}^{2} b^{2}_{1} }{\hslash^{2}}  \right) \
\nonumber
\\
&&
\times
L^{|m-\bar{m}|}_{n_{2} + min(m,\bar{m})} \left(\frac{2q_{2}^{2}}{b^{2}_{2}} + \frac{2p_{1}^{2} b^{2}_{2} }{\hslash^{2}}  \right),
\label{wignerFunctionThermoBellCatState}
\end{eqnarray}
where we defined the constants $\displaystyle b_{i}^{2}=\frac{\hslash}{m \omega_{i}}$ and
$\displaystyle \chi_{i} \ = \ \left\lbrace \begin{array}{l}
\displaystyle \frac{q_{i}}{b_{i}} + i\frac{p_{i}b_{i}}{\hslash}, \ \ \textrm{ if } \ n \geq \bar{n} \\
\displaystyle \frac{q_{i}}{b_{i}} - i\frac{p_{i}b_{i}}{\hslash}, \ \ \textrm{ if } \ n < \bar{n} 
\end{array} \right. $ with $i=\lbrace 1,2 \rbrace$.

\begin{figure}[!ht]
\centering 

\vspace{0.8cm} 
\hspace{-5.0cm}
\subfigure[State $\vert \Psi_{+}(\beta) \rangle$ with  $T=0.01\textrm{K}$]{
\includegraphics[width=0.45\textwidth,trim= 20mm 25mm 10mm 8mm,clip]{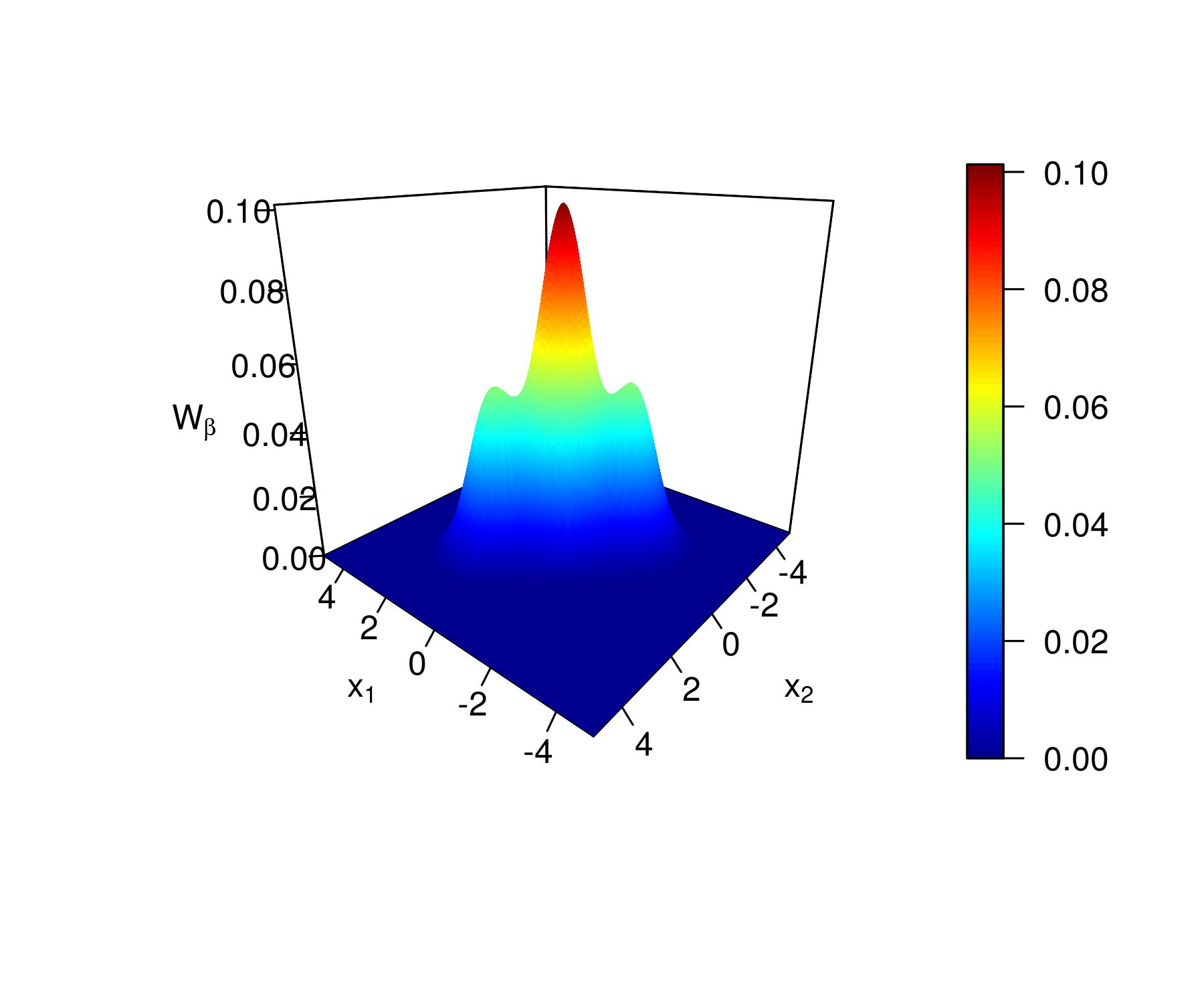} }
\hspace{0.5cm}
\subfigure[ State $\vert \Phi_{-}(\beta) \rangle$ with  $T=0.01\textrm{K}$]{
\includegraphics[width=0.45\textwidth,trim= 20mm 25mm 10mm 0mm,clip]{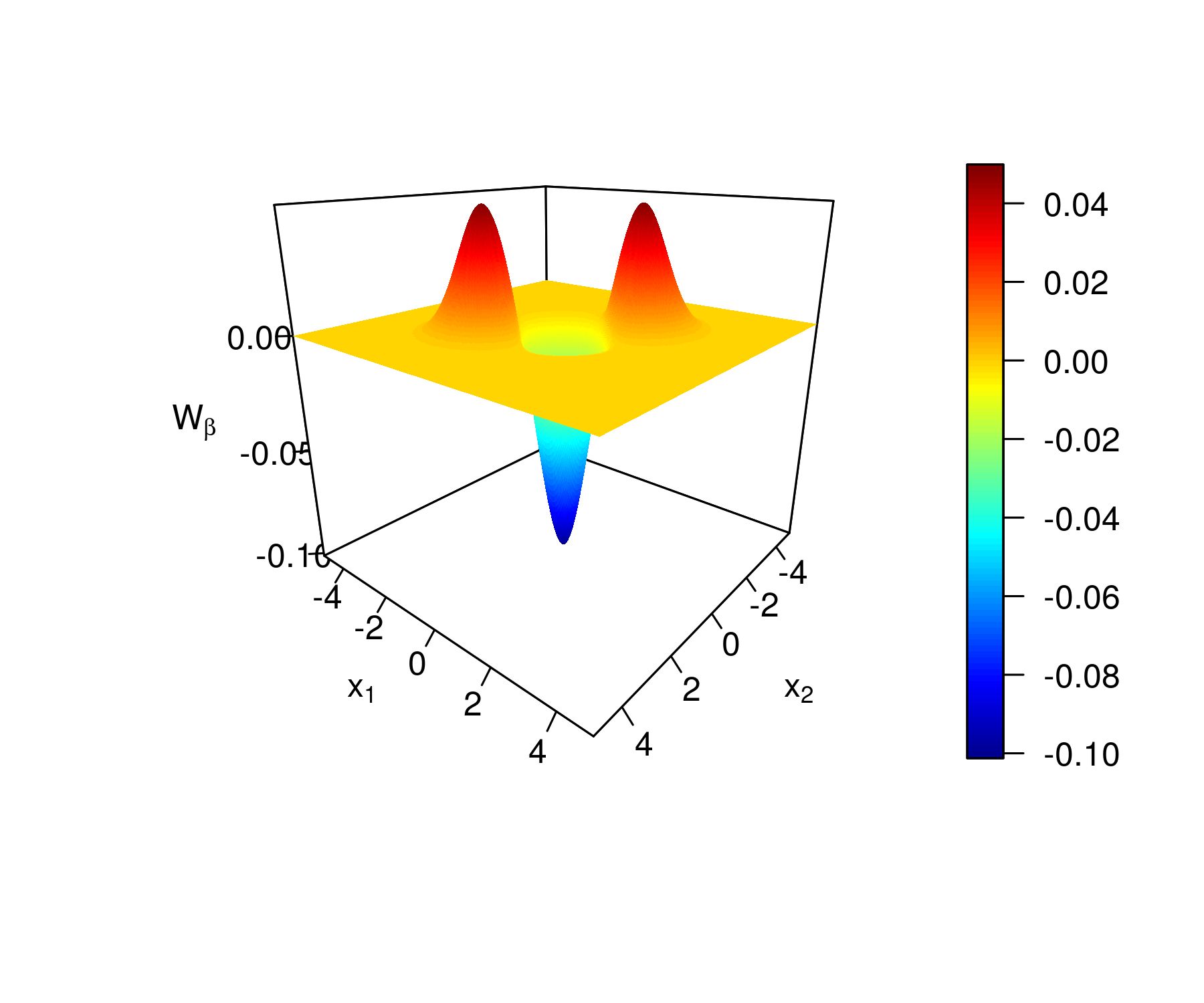}  }\hspace{-5.5cm}

\hspace{-5.0cm} \subfigure[State $\vert \Psi_{+}(\beta) \rangle$ with  $T=1 \textrm{K}$]{
 \includegraphics[width=0.45\textwidth,trim= 20mm 25mm 10mm 0mm,clip]{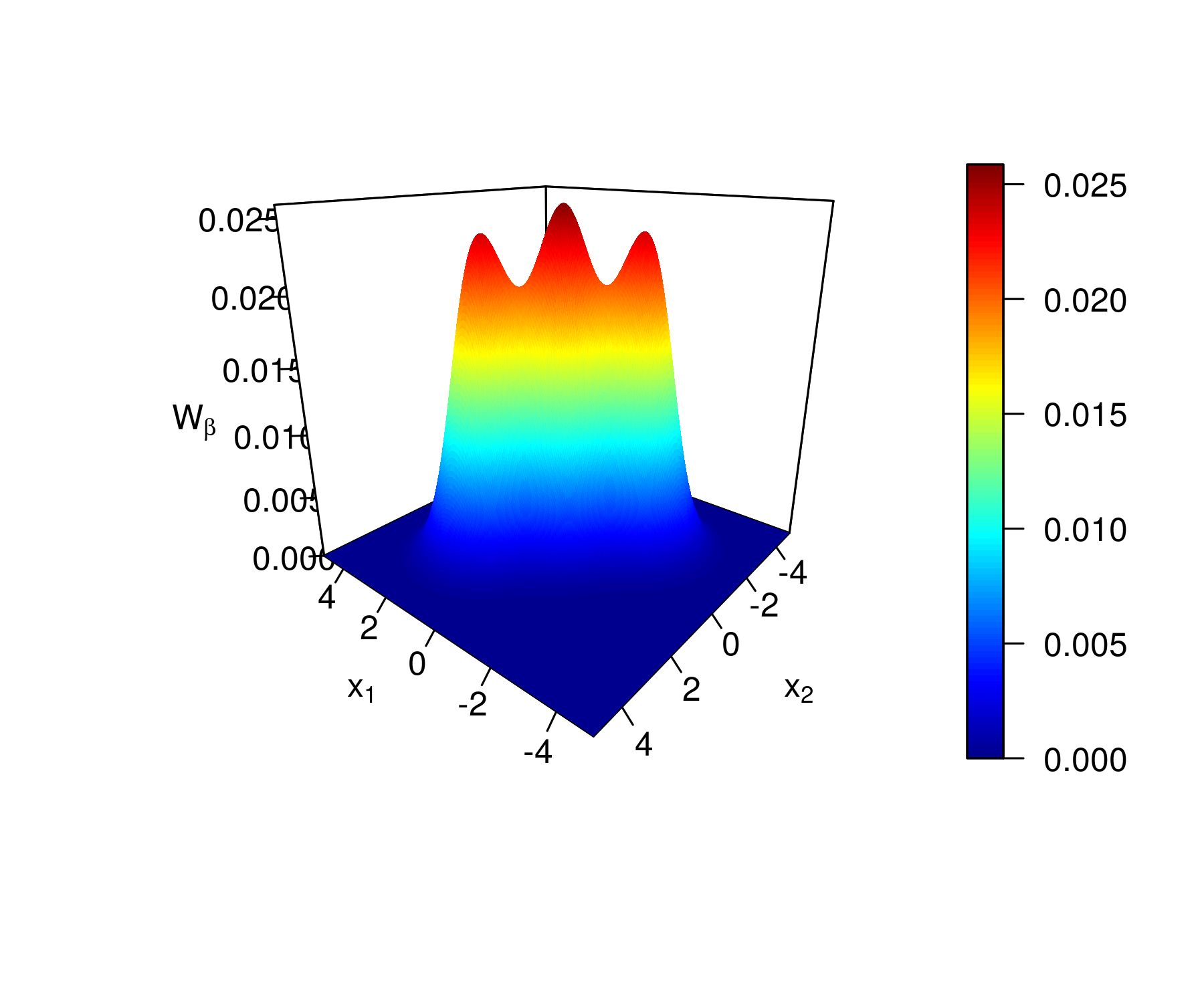}
}
\hspace{0.5cm}
 \subfigure[State $\vert \Phi_{-}(\beta) \rangle$ with  $T=1 \textrm{K}$]{ 
 \includegraphics[width=0.45\textwidth,trim= 20mm 25mm 10mm 0mm,clip]{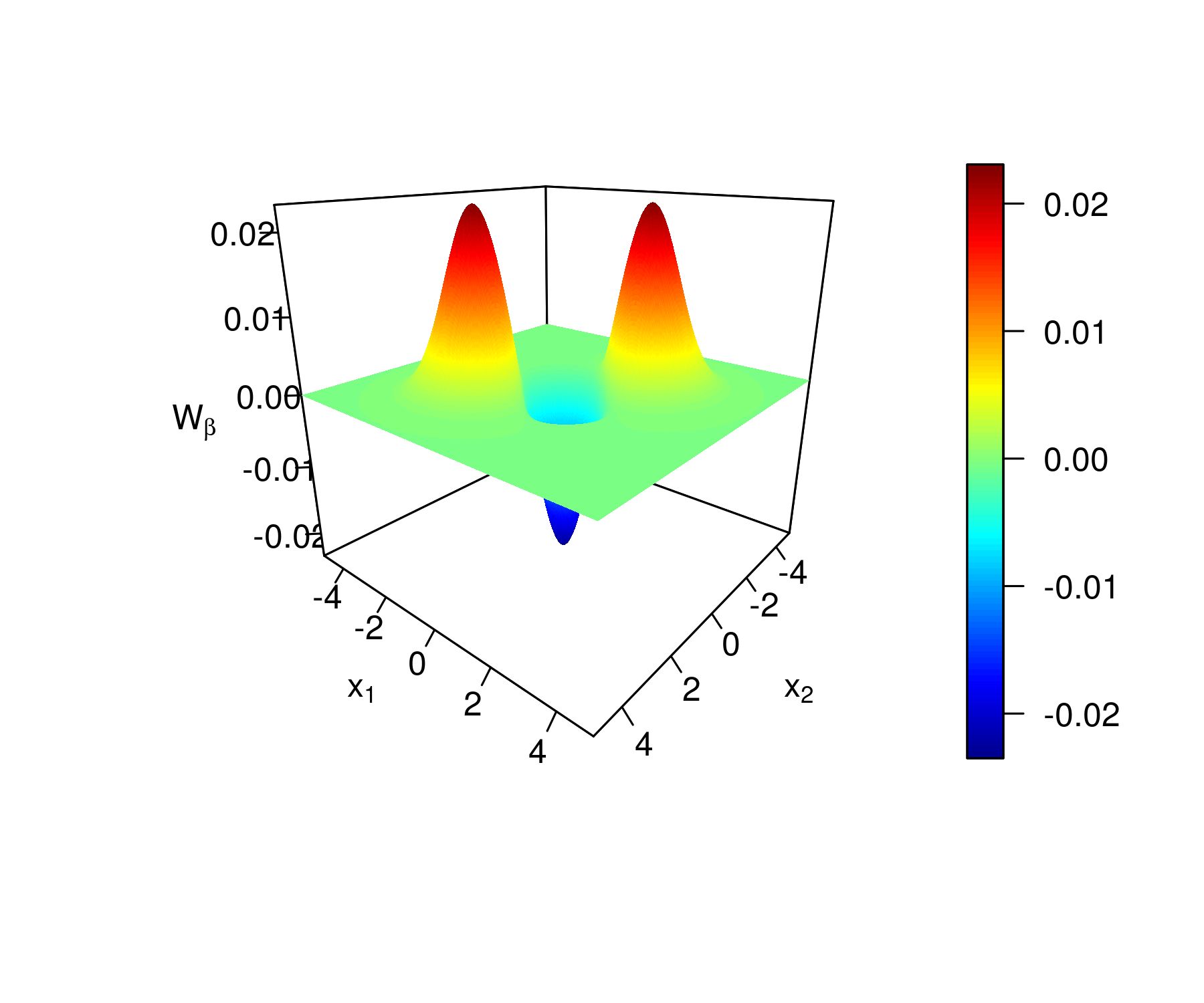} 
} \hspace{-5.5cm}

\hspace{-5.0cm} \subfigure[State $\vert \Psi_{+}(\beta) \rangle$ with  $T=10\textrm{K}$]{
 \includegraphics[width=0.45\textwidth,trim= 20mm 25mm 10mm 0mm,clip]{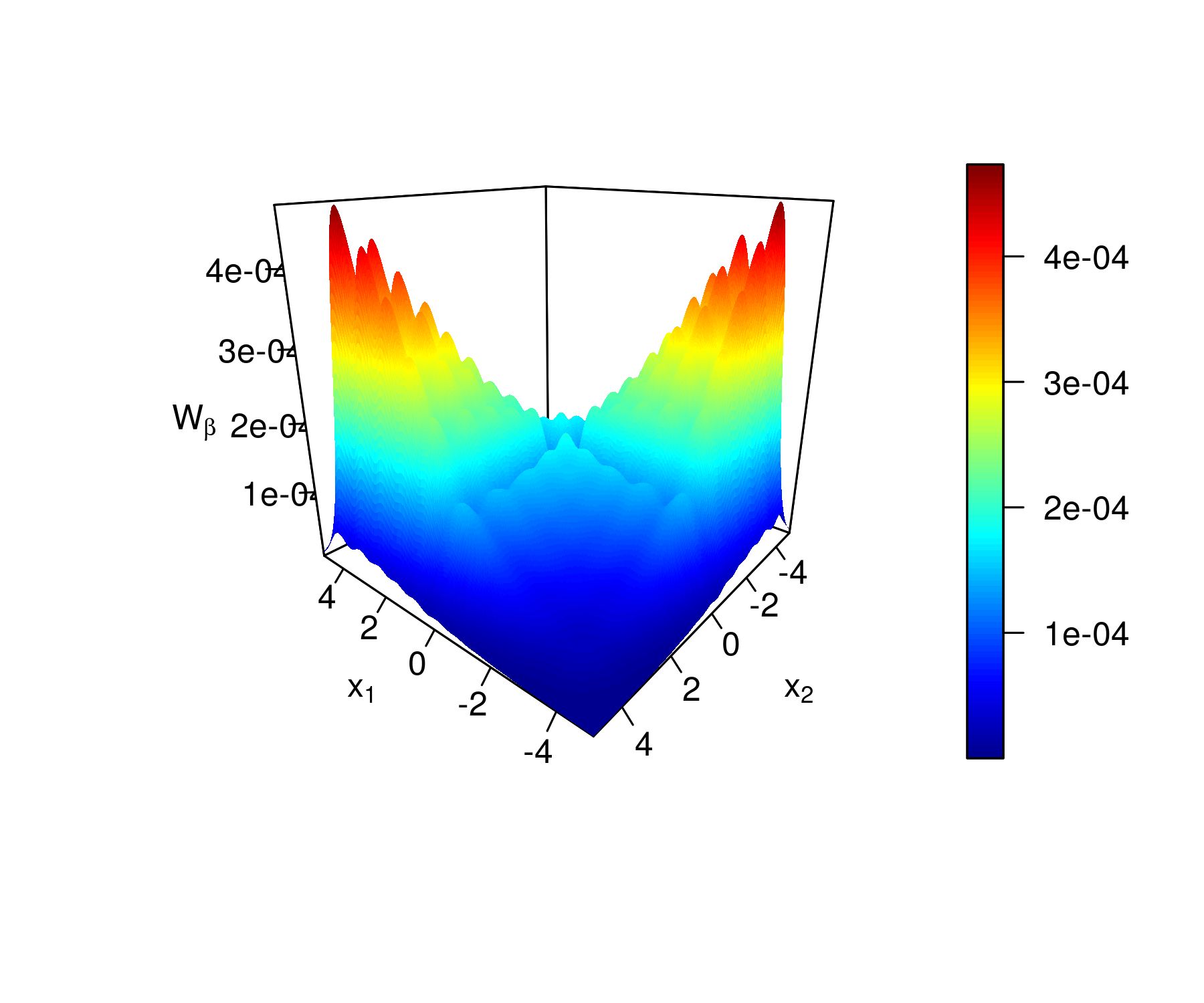} 
}
\hspace{0.5cm}
\subfigure[State $\vert \Phi_{-}(\beta) \rangle$ with  $T=10\textrm{K}$]{
 \includegraphics[width=0.45\textwidth,trim= 20mm 25mm 10mm 0mm,clip]{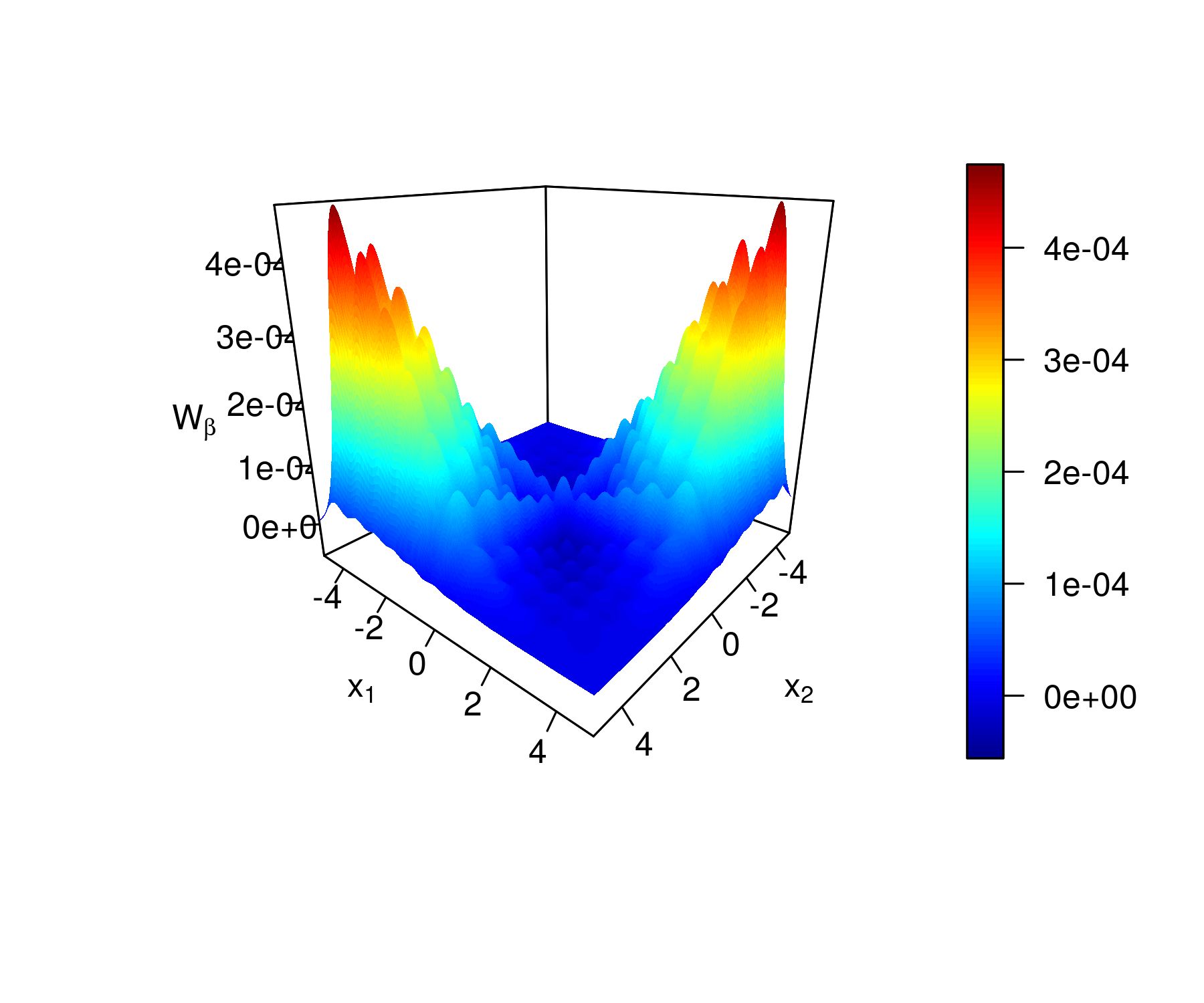} 
} \hspace{-5.5cm} 
\caption{Wigner function of thermal Bell-Cat states with $\alpha= 1$.} \vspace{-3.5cm}

\label{fig-alpha1}
\end{figure}

\begin{figure}[!ht]
\centering

\vspace{0.8cm} 
\hspace{-5.0cm} \subfigure[State $\vert \Psi_{+}(\beta) \rangle$ with  $T=0.01\textrm{K}$]{
\includegraphics[width=0.45\textwidth,trim= 20mm 25mm 10mm 10mm,clip]{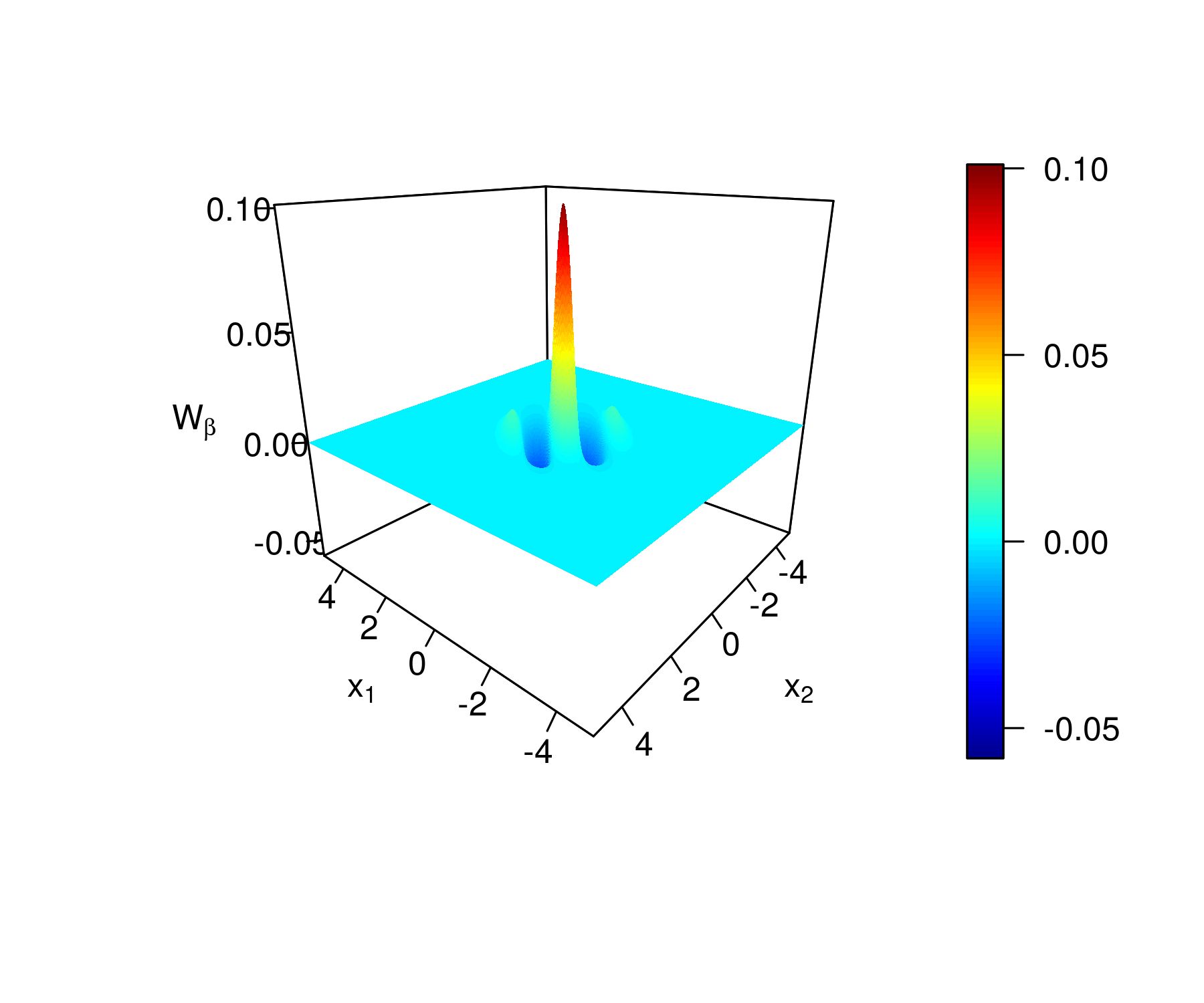}
}
\hspace{0.5cm}
\subfigure[ State $\vert \Phi_{-}(\beta) \rangle$ with  $T=0.01\textrm{K}$]{
\includegraphics[width=0.45\textwidth,trim= 20mm 25mm 10mm 0mm,clip]{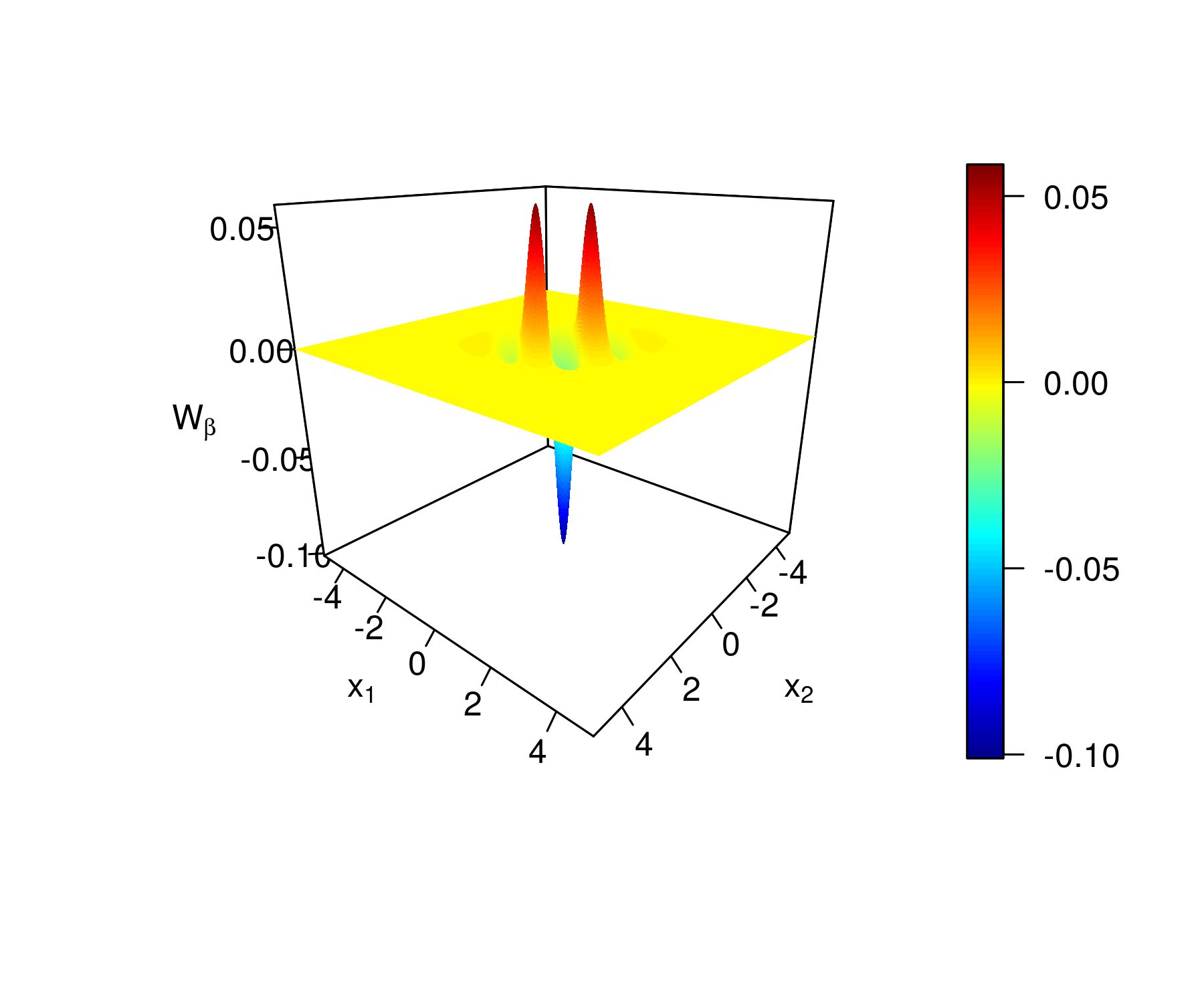} 
} \hspace{-5.5cm}

\hspace{-5.0cm} \subfigure[State $\vert \Psi_{+}(\beta) \rangle$ with  $T=1 \textrm{K}$]{
\includegraphics[width=0.45\textwidth,trim= 20mm 25mm 10mm 0mm,clip]{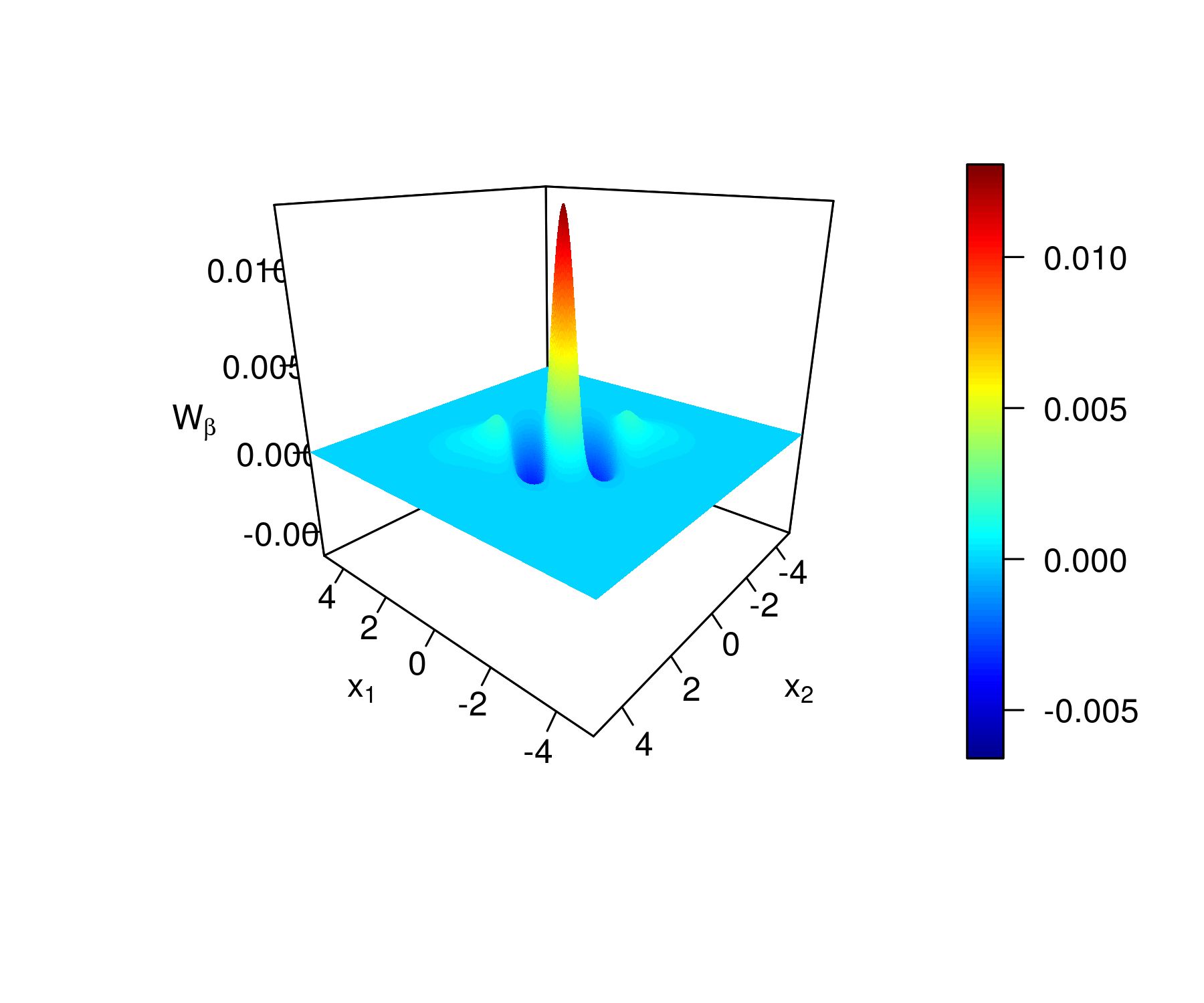} 
}
\hspace{0.5cm}
\subfigure[State $\vert \Phi_{-}(\beta) \rangle$ with  $T=1 \textrm{K}$]{
\includegraphics[width=0.45\textwidth,trim= 20mm 25mm 10mm 0mm,clip]{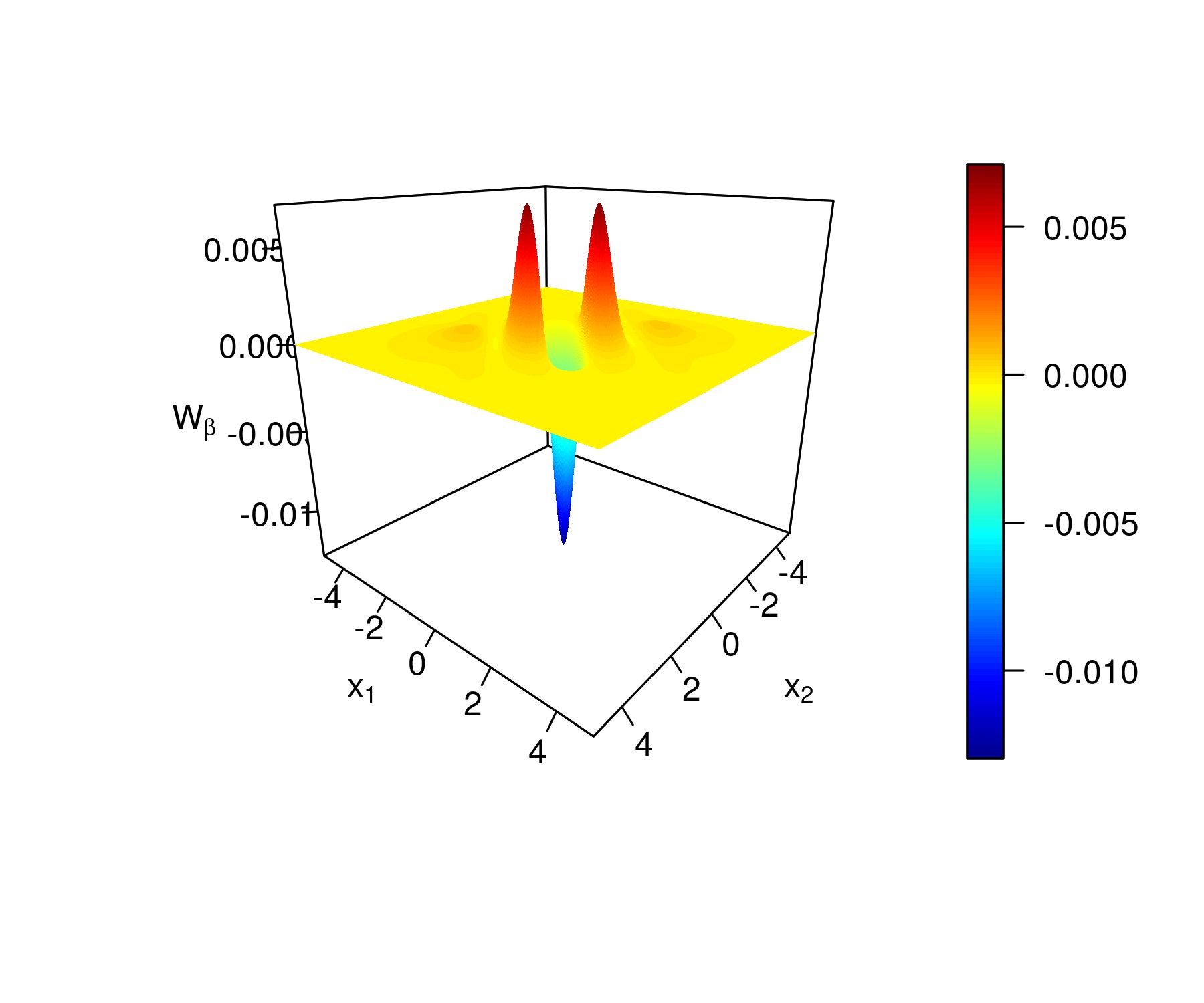} 
}\hspace{-5.5cm}

\hspace{-5.0cm} \subfigure[State $\vert \Psi_{+}(\beta) \rangle$ with  $T=10\textrm{K}$]{
\includegraphics[width=0.45\textwidth,trim= 20mm 25mm 10mm 0mm,clip]{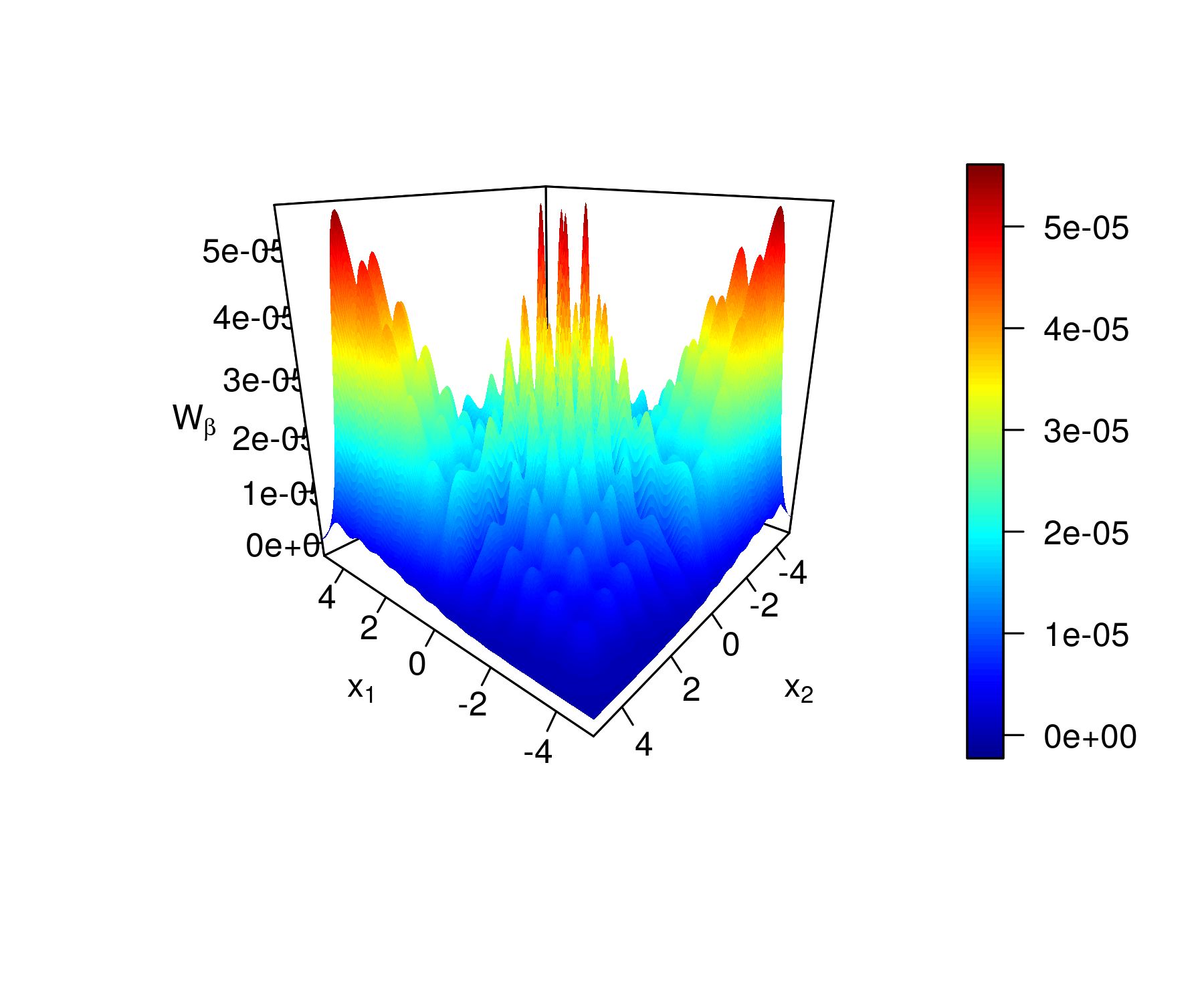}
}
\hspace{0.5cm}
\subfigure[State $\vert \Phi_{-}(\beta) \rangle$ with  $T=10\textrm{K}$]{
 \includegraphics[width=0.45\textwidth,trim= 20mm 25mm 10mm 0mm,clip]{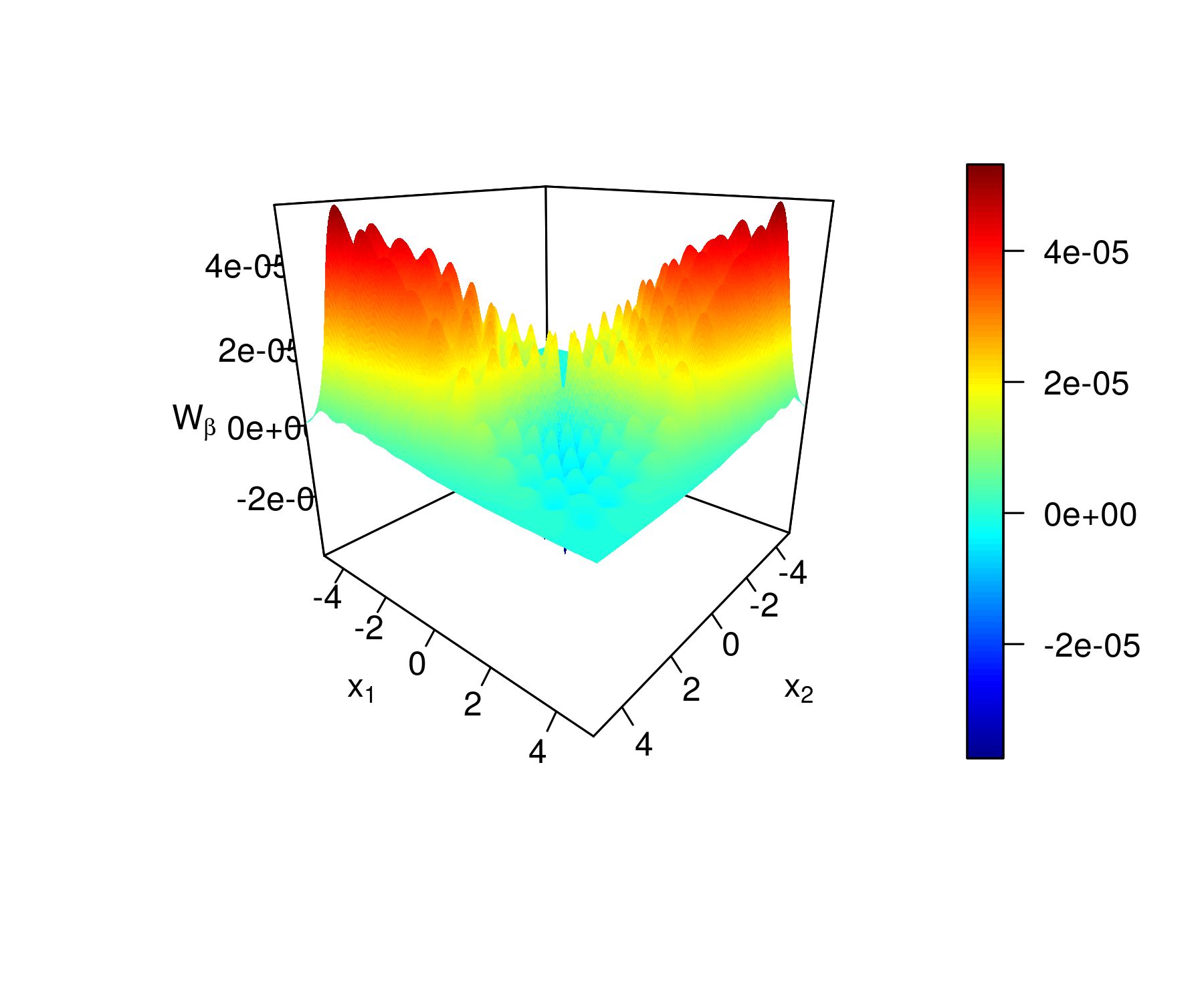}
 } \hspace{-5.5cm} 
\caption{Wigner function of thermal Bell-Cat states with $\alpha= 1 + i$.}   
\vspace{-3.5cm}
\label{fig-alpha1i}
\end{figure}

\begin{figure}[!ht]
\centering

\vspace{0.8cm} 
\hspace{-5.0cm} \subfigure[State $\vert \Psi_{+}(\beta) \rangle$ with  $T=0.01\textrm{K}$]{
\includegraphics[width=0.45\textwidth,trim= 20mm 25mm 10mm 8mm,clip]{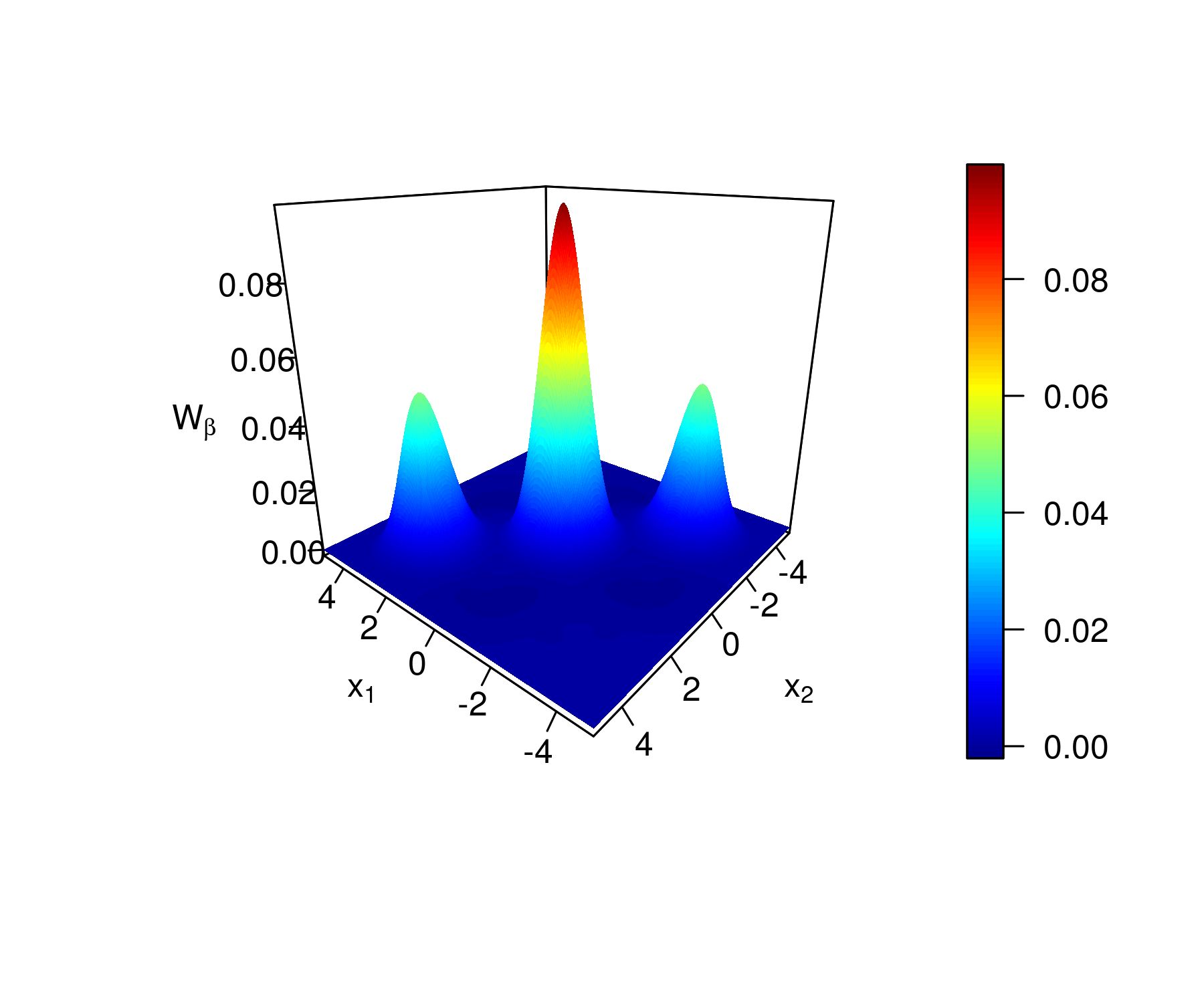}
}
\hspace{0.5cm}
\subfigure[State $\vert \Phi_{-}(\beta) \rangle$ with  $T=0.01\textrm{K}$]{
\includegraphics[width=0.45\textwidth,trim= 20mm 25mm 10mm 0mm,clip]{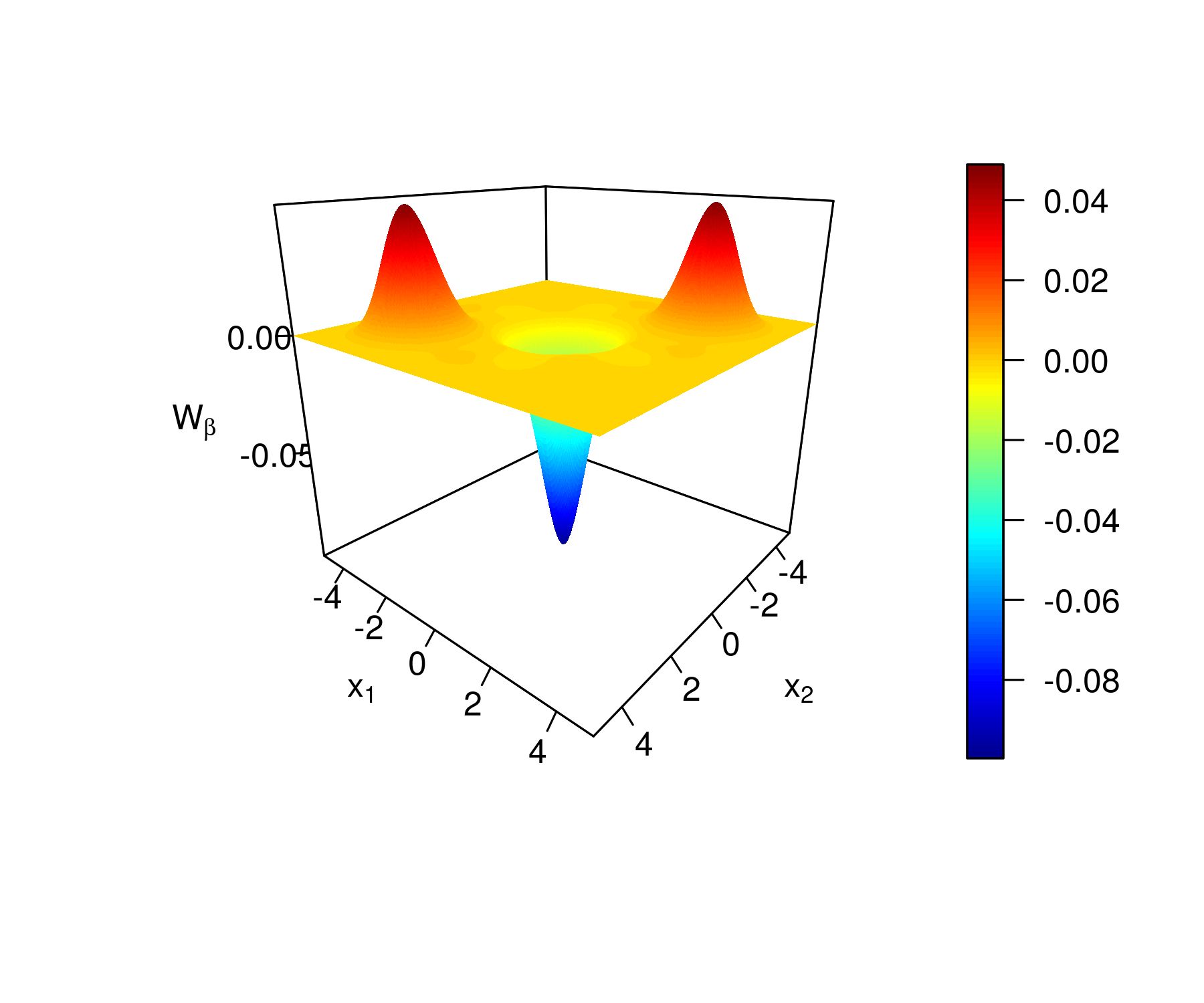}
}\hspace{-5.5cm}

\hspace{-5.0cm}\subfigure[State $\vert \Psi_{+}(\beta) \rangle$ with  $T=1 \textrm{K}$]{
\includegraphics[width=0.45\textwidth,trim= 20mm 25mm 10mm 0mm,clip]{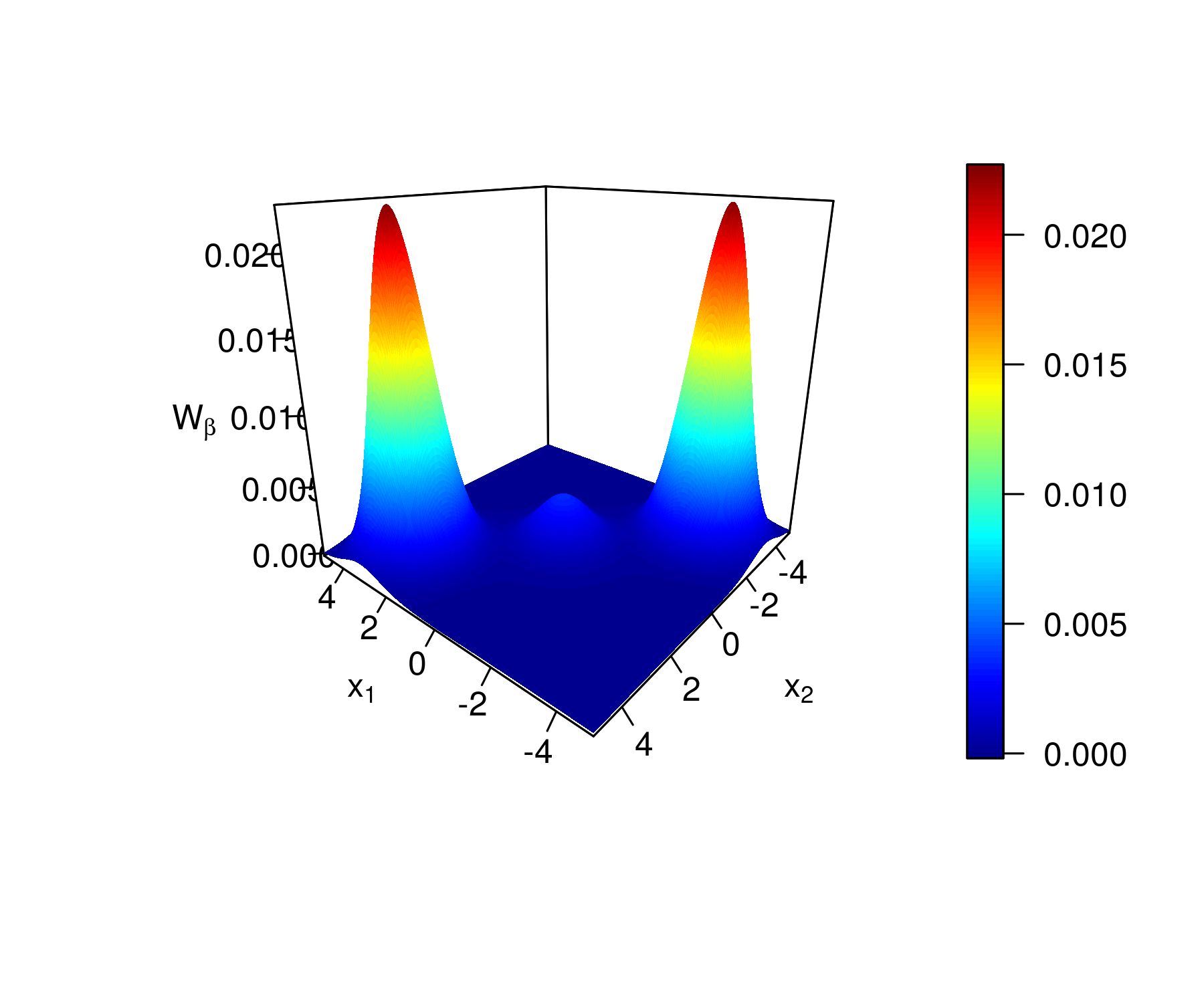} 
}
\hspace{0.5cm}
\subfigure[State $\vert \Phi_{-}(\beta) \rangle$ with  $T=1 \textrm{K}$]{
\includegraphics[width=0.45\textwidth,trim= 20mm 25mm 10mm 0mm,clip]{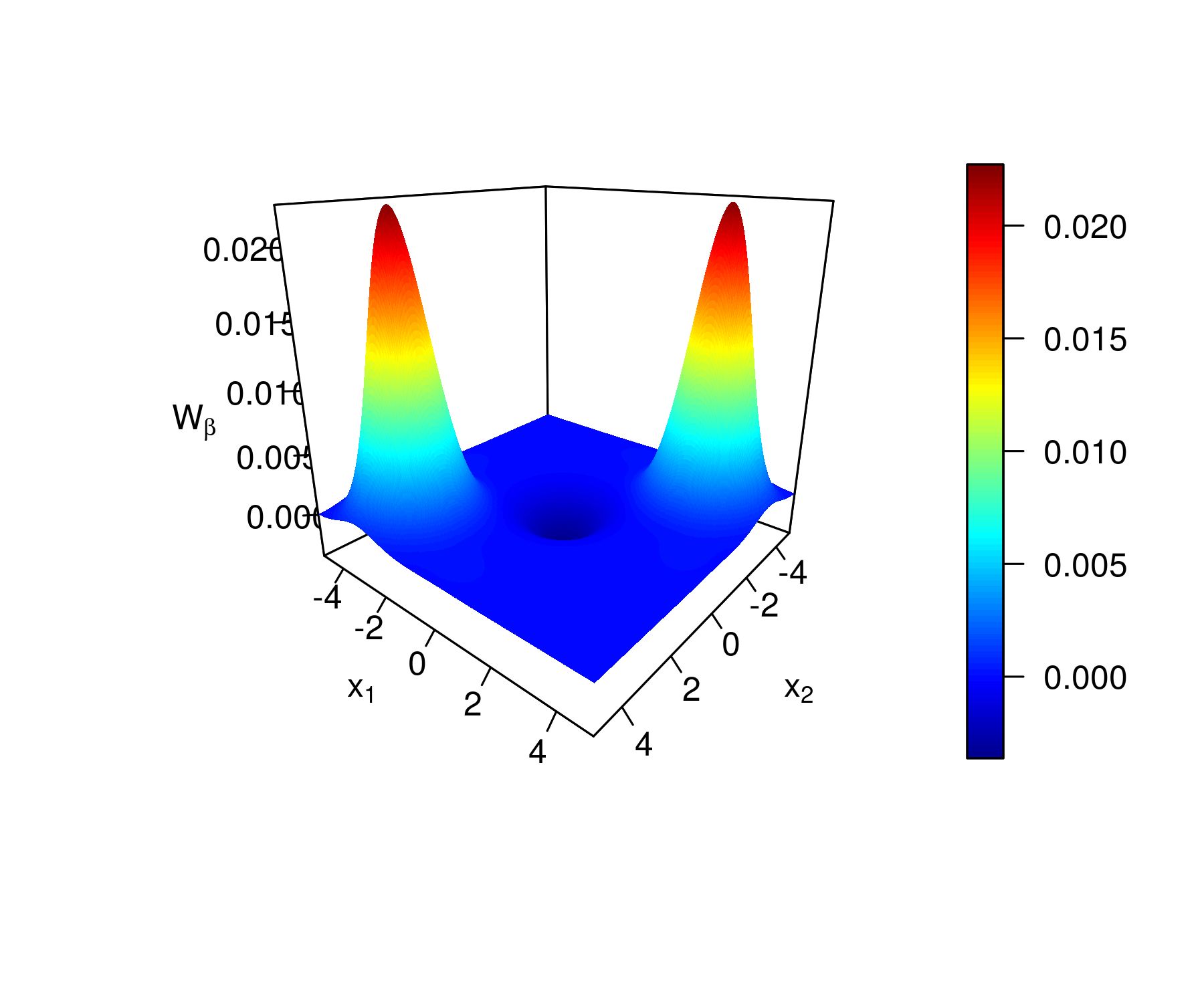}
} \hspace{-5.5cm}

\hspace{-5.0cm}\subfigure[State $\vert \Psi_{+}(\beta) \rangle$ with  $T=10\textrm{K}$]{
\includegraphics[width=0.45\textwidth,trim= 20mm 25mm 10mm 0mm,clip]{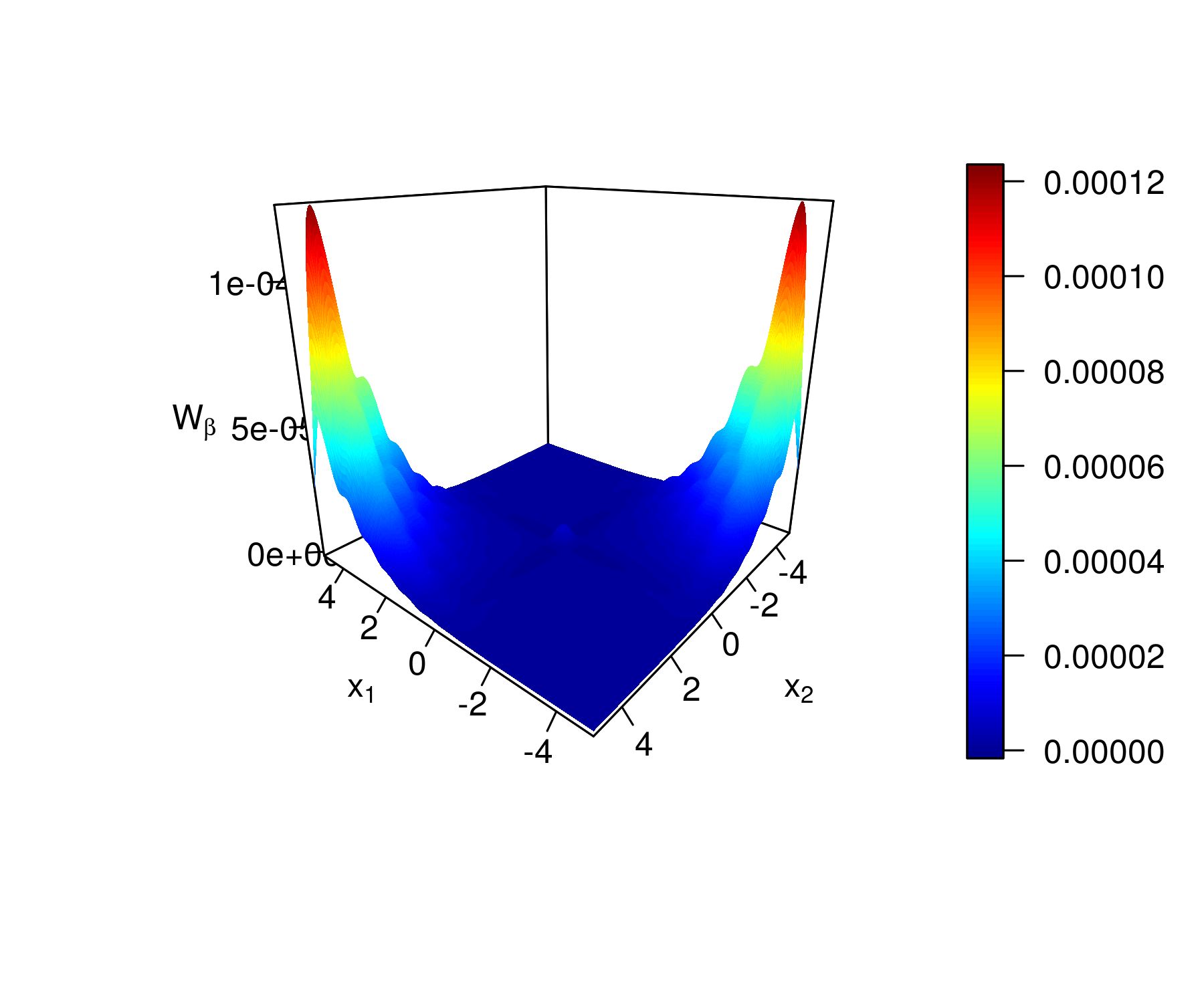}
}
\hspace{0.5cm}
\subfigure[State $\vert \Phi_{-}(\beta) \rangle$ with  $T=10\textrm{K}$]{
\includegraphics[width=0.45\textwidth,trim= 20mm 25mm 10mm 0mm,clip]{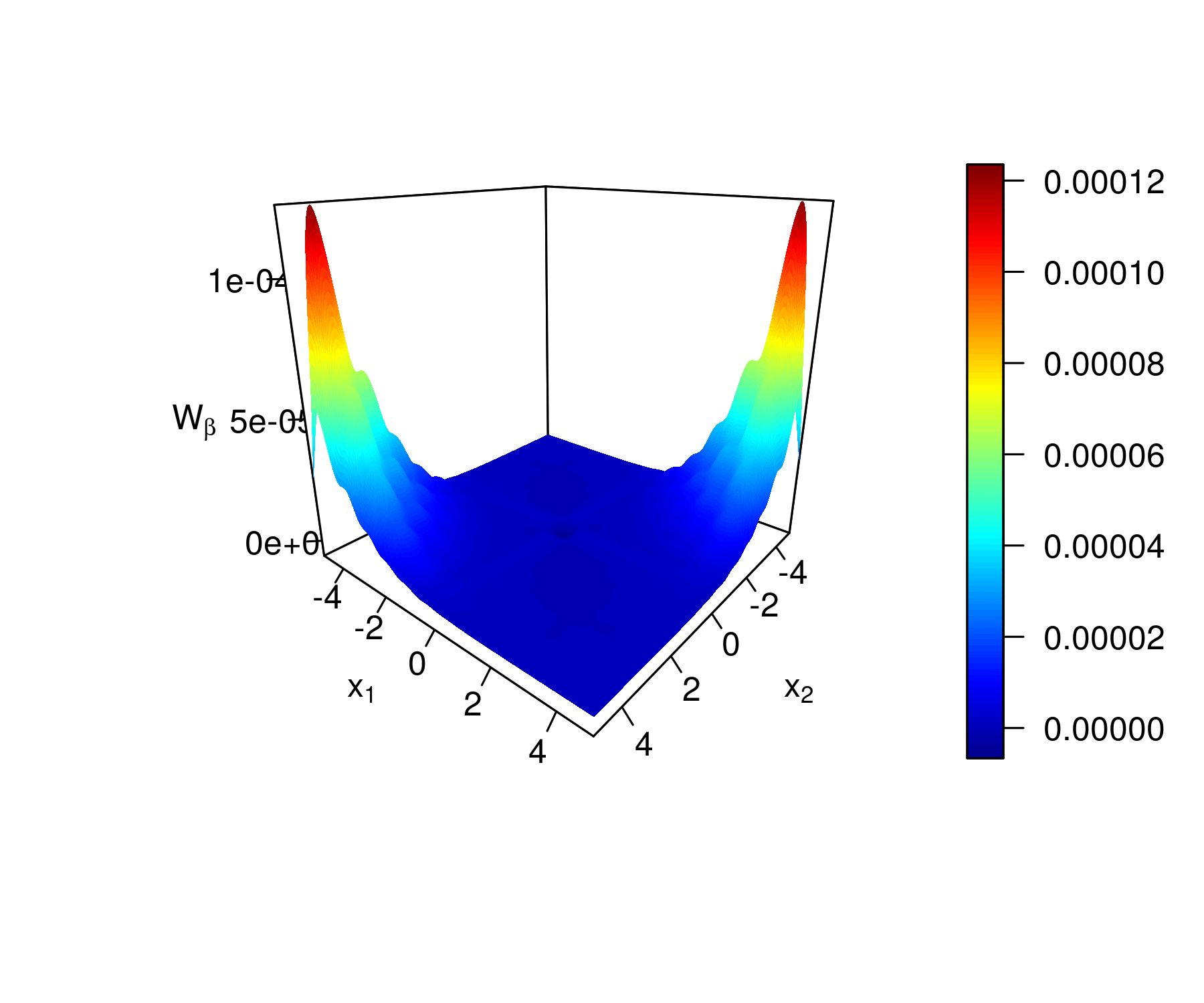} 
}\hspace{-5.5cm} 
\caption{Wigner function of thermal Bell-Cat states with $\alpha= 2$.} 
\vspace{-3.5cm}
\label{fig-alpha2}
\end{figure}

The interest in obtaining the WF of the quantum states used in quantum optics is related to its negativity properties. 
Our Wigner functions are  functions in four dimensions and  cannot be completely visualized in a three dimensional plot unless we specify values for $q_{i}$ and $p_{i}$. This can be done in several ways. In figure \ref{fig-alpha1}, \ref{fig-alpha1i}  and \ref{fig-alpha2} $ \ $ we $ \ $ choose $ \ $ to $ \ $ represent $\displaystyle x_{1} = \frac{q_{1}}{b_{1}} + i\frac{p_{1}b_{1}}{\hslash}$
and $\displaystyle x_{2} = \frac{q_{2}}{b_{2}} + i\frac{p_{2}b_{2}}{\hslash}$  so that we can visualize the $ \ \displaystyle W_{\alpha,\beta}^{k,\pm}(x_{1},x_{2}) \ $ portion of the complete WF. Based $ \ $  in  $ \ $  the 
Refs. \cite{wang2016,vla2015,song2019} we choose $\displaystyle \frac{\omega_{1}}{2\pi} = \frac{\omega_{2}}{2\pi} = 5.5 \:\textrm{GHz}$ and performed the calculation of the function (\ref{wignerFunctionThermoBellCatState}) numerically using the R Language \cite{rcran}. The results are presented in the figures  \ref{fig-alpha1}, \ref{fig-alpha1i}  and \ref{fig-alpha2}.

These figures show the plots of the WF associated with the states $\vert \Psi_{+}(\beta)\rangle$ and $\vert \Phi_{-}(\beta)\rangle$ with the parameter $\alpha$ possessing the values: $\alpha= 1$, $\alpha= 1+i$ and $\alpha= 2$, and temperature values: $T=0.01\textrm{K}$, $T=1\textrm{K}$ and $T=10\textrm{K}$. We choose only the states $\vert \Psi_{+}(\beta)\rangle$ and $\vert \Phi_{-}(\beta)\rangle$ in our analysis, because these states correspond to a rotation in the phase space of the states  $\vert \Phi_{+}(\beta)\rangle$ and $\vert \Psi_{-}(\beta)\rangle$ respectively. Our graphics coincide with those obtained in Ref. \cite{weinbub2018} for lower temperature, reinforcing the results.

\newpage


\newpage


\newpage


\newpage

Varying the temperature we can see in more datails the importance of thermal effects in the analysis of Bell-Cat states. 
In figures  \ref{fig-alpha1},  \ref{fig-alpha1i} and  \ref{fig-alpha2} we observe a decrease in the range of both positive and negative values of the thermal WF as the temperature goes higher. For temperatures smaller than $0.01 \textrm{K}$ we have a similar behavior, which indicates that thermal effects have less influence in these temperatures. As the temperature increases, such for example at $10\textrm{K}$, the Bell-Cat states loose their properties. This is suggested by the deformed shape of the thermal WF graphics showed in the subfigures $(e)$ and $(f)$ of the figures \ref{fig-alpha1},  \ref{fig-alpha1i} and  \ref{fig-alpha2}.

The restriction of the plot of the WF, corresponding to Bell-Cat states, to three dimentions is not the best method to analyze the non-classicality, because these plots are particular situations and consequently do not show all information. Another way to analyze the behaviour of the WF is to use the volume of the negative part. We highlight two measures of negavity of WF and apply to our thermal expression (\ref{wignerFunctionThermoBellCatState}). The first one was proposed as the phase space integral of the difference between the absolute value and the actual value of the WF \cite{kenfack2004}, defined as
\begin{eqnarray}
\hspace{-1cm}
\delta_{\alpha,\beta}^{k,\pm}
& = &
\int\!\!\! \int\!\!\! \int\!\!\! \int 
\left[ 
 \ |W_{\alpha,\beta}^{k,\pm}(q_{1},p_{1},q_{2},p_{2})|  - W_{\alpha,\beta}^{k,\pm}(q_{1},p_{1},q_{2},p_{2}) 
\right] dq_{1} dp_{1} dq_{2} dp_{2} \label{delta-1}.
\end{eqnarray}

The second one is based on the rate of the sum and difference between the positives and negatives volumes of the WF \cite{benedict99,foldi2002}, which when applied to our expression yields
\begin{eqnarray}
\nu_{\alpha,\beta}^{k,\pm}
& =  &
 1 -\frac{I_{+}(W_{\alpha,\beta}^{k,\pm}) - I_{-}(W_{\alpha,\beta}^{k,\pm}) }{ I_{+}(W_{\alpha,\beta}^{k,\pm}) + I_{-}(W_{\alpha,\beta}^{k,\pm})} \label{nu-1},
\end{eqnarray}
where $I_{+}(W_{\alpha,\beta}^{k,\pm})$ and $I_{-}(W_{\alpha,\beta}^{k,\pm})$ are the volume modules of positive and negative parts of the thermal Wigner functions (\ref{wignerFunctionThermoBellCatState}), respectively. If we consider the normalization $I_{+}(W_{\alpha,\beta}^{k,\pm}) - I_{-}(W_{\alpha,\beta}^{k,\pm}) = 1$ we can relate these parameters through the following expression:
\begin{eqnarray}
\nu_{\alpha,\beta}^{k,\pm} 
& = &
\frac{2 I_{-}(W_{\alpha,\beta}^{k,\pm})}{1+ 2I_{-}(W_{\alpha,\beta}^{k,\pm})} \ = \  \frac{\delta_{\alpha,\beta}^{k,\pm}}{1+ \delta_{\alpha,\beta}^{k,\pm}}.
\label{nuBeta}
\end{eqnarray}

\begin{figure}[!ht]
\centering

 \hspace{-5.5cm}
\subfigure[State $\displaystyle \vert \Psi_{+}(\beta) \rangle$ ]{
 \includegraphics[scale=0.6,trim= 0mm 0mm 0mm 10mm,clip]{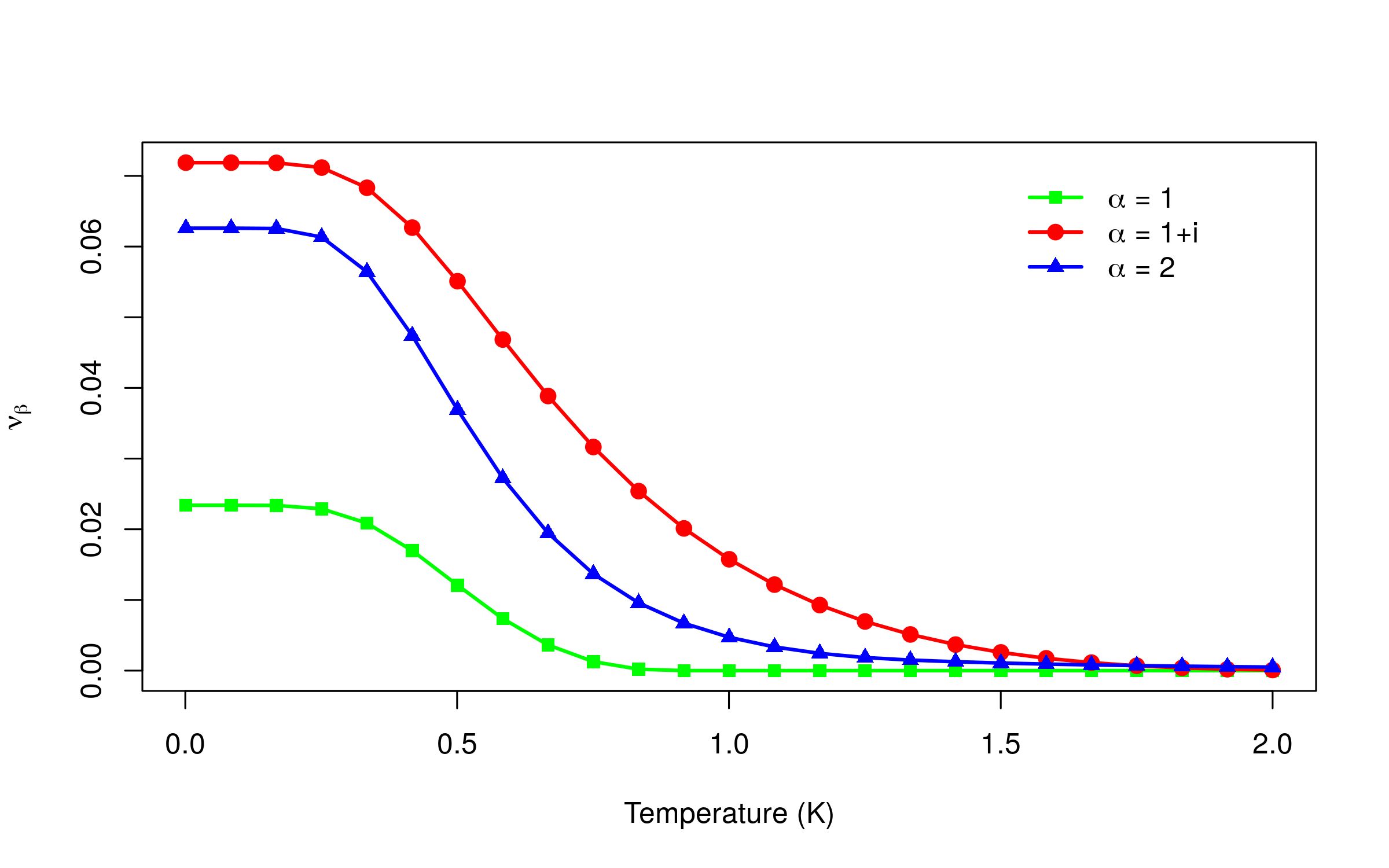}  }
 \hspace{-6.0cm}

\vspace{-0.35cm} \hspace{-5.5cm}
\subfigure[State $\displaystyle \vert \Phi_{-}(\beta) \rangle$ ]{ 
\includegraphics[scale=0.6]{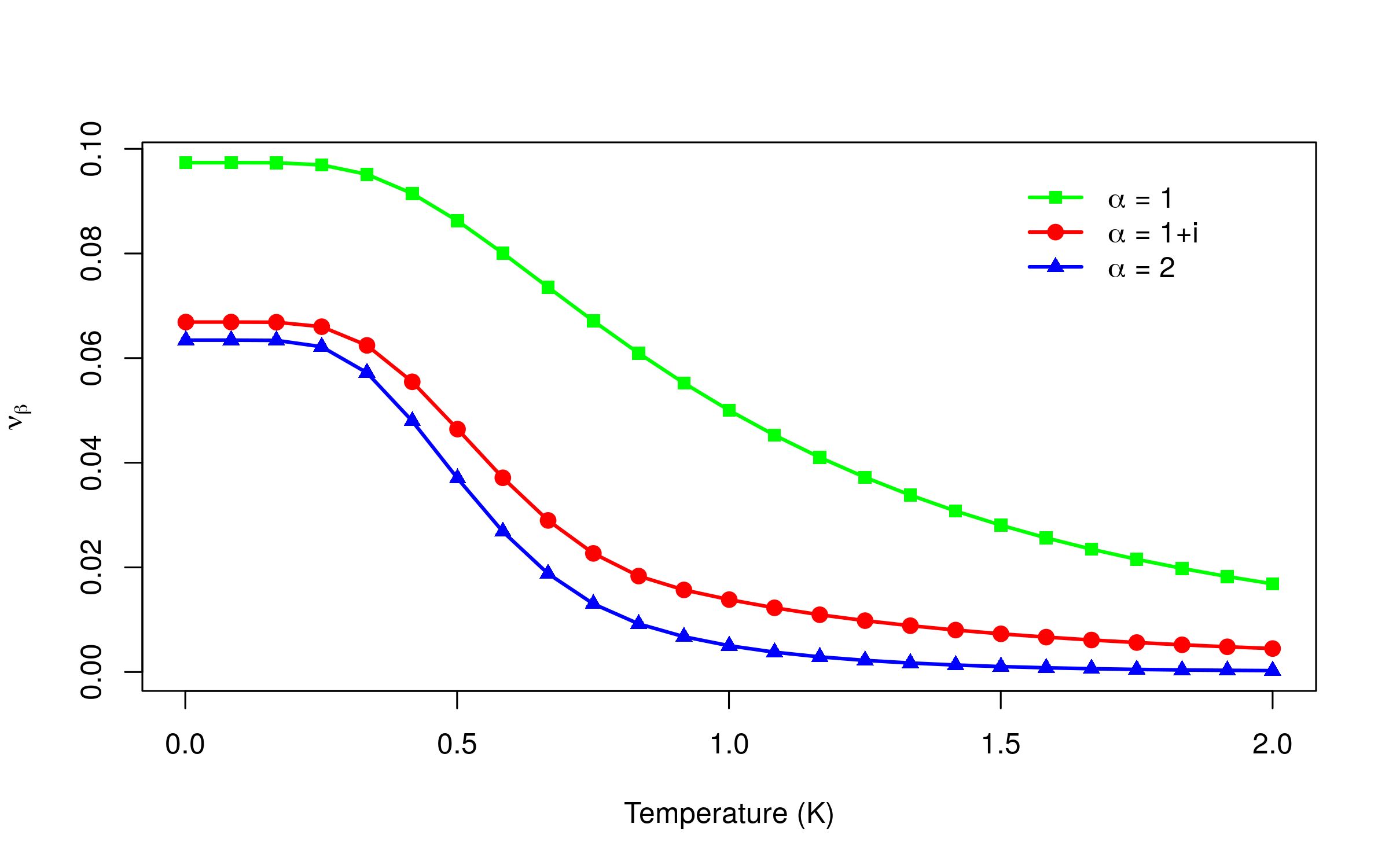}  } \hspace{-6.0cm}

\caption{Negativity of Thermal Wigner function of thermal Bell-Cat states with $\displaystyle \alpha= \lbrace 1, 1+i, 2 \rbrace$.}
\label{fig-nu}
\end{figure}

We plot in figure \ref{fig-nu} the curve of the $\nu_{\alpha,\beta}^{k,\pm}$ parameter for several temperature values and visualize how non-classicality evolves towards gradual changes of thermal equilibrium of the system. Numerical integration of the thermal WF (\ref{wignerFunctionThermoBellCatState}) was performed by R Language \cite{rcran}, for temperature values varying from $0\textrm{K}$ to $2\textrm{K}$ and was used to calculate the parameter $\nu_{\alpha,\beta}^{k,\pm} $ corresponding to the states  $\vert \Psi_{+} (\beta)\rangle$ and $\vert \Phi_{-}(\beta)\rangle$, with $\alpha= 1$, $\alpha= 1+i$ and $\alpha= 2$. We can observe that the parameter $\nu_{\alpha,\beta}^{k,\pm} $ measuring non-classicality remains stable until approximately  $T \approx 0.3\textrm{K}$, which coincides with the values found in the Refs. \cite{kenfack2004,alexei}.  From this temperature onwards $\nu_{\alpha,\beta}^{k,\pm} $ presents a gradual decrease getting closer to zero as the temperature grows. For the state $\vert \Psi_{+} (\beta)\rangle$ the highest deviation to non-classical properties is reached for $\alpha=1+i$ whereas for $\vert \Phi_{-}(\beta)\rangle$ the highest deviation is reached for $\alpha=1$ exhibiting the largest of the non-classical properties among all we have analyzed.  These properties remain over a larger temperature range.

Despite the limitations of three dimensional plot analysis of a four dimensional function, the behavior discussed above could be read from the figures \ref{fig-alpha1},  \ref{fig-alpha1i} and  \ref{fig-alpha2} where the changing in the peaks of the Wigner functions is observed as the value of temperature increases from $T=0.01\textrm{K}$ to $T=10\textrm{K}$. We further note the decrease of the negative peaks, which carry the non-classicality of the states, for the higher temperature.

\section{Conclusions \label{sec5}}

In this paper we have used the framework of Thermofield Dynamics in order to include temperature in the study of Bell-Cat states. We have found the analytical expressions for the thermal density operators as well as the thermal Wigner functions associated with these states. The analyses of the three dimensional graphics of the thermal Wigner functions associated with the states  $  \vert \Psi_{+} (\beta)\rangle$ and $\vert \Phi_{-}(\beta)\rangle  $, with $\alpha= 1$, $\alpha= 1+i$ and $\alpha= 2$, and for the temperature of  $T=\lbrace 0.01\textrm{K}, 1\textrm{K}, 10\textrm{K} \rbrace$, suggest that, for an increase in temperature, there is a decrease in the values of the thermal Wigner functions. Such behavior can be interpreted as a gradual loss in its properties as temperature increases.
To complement our analysis we proposed to use the negativity parameter idealized by Benedict \textit{et al.} \cite{benedict99,foldi2002} to investigate the non-classicality of the thermal Wigner functions associated with the two modes Bell-Cat states.  This concept applied to thermal Wigner functions results in a temperature dependent parameter $\nu_{\alpha,\beta}^{k,\pm}$  for the temperature range made explicit above. The $\nu_{\alpha,\beta}^{k,\pm}$ parameter shows that for temperatures higher than $0.3\textrm{K}$ the Bell-Cat states gradually loose their non-classical properties until around $2\textrm{K}$, where the read of the $\nu_{\alpha,\beta}^{k,\pm}$ suggests that these properties subside.  
Combining  Benedict's parameter with the thermal Wigner function seems to be promising strategy since the non-classical properties can be associated with the negativity of the volume associated with the Wigner function. This proved helpful in the analysis of the non-classicality behaviour with respect to the temperature. This analysis can be further applied to states such as those of interest in quantum optics and quantum information. In addition this method can be extended by looking for analytical expression for the limits where non-classical properties of the systems remain preserved.

\ack

This study was financed in part by the Coordena\c{c}\~ao de Aperfei\c{c}oamento de Pessoal de N\'ivel Superior - Brasil (CAPES) - Finance Code 001.


\section*{References}


\begin{thebibliography}{99}
\bibitem{EPR}{   Einstein A,  Podolsky B and  Rosen N 1935 {\it Phys. Rev. } \href{https://journals.aps.org/pr/abstract/10.1103/PhysRev.47.777}{ {\bf 47} 777  }}


\bibitem{Bell}{   Bell J S 1964 {\it Physics Physique Fizika} \href{https://doi.org/10.1103/PhysicsPhysiqueFizika.1.195}{ {\bf 1} 195  }}


\bibitem{Aspect}{ Aspect A,  Grangier P and  Rogers G 1982 {\it Phys. Rev. Lett.} \href{https://journals.aps.org/prl/abstract/10.1103/PhysRevLett.49.91}{ {\bf 48} 91};
   Aspect A,  Dalibard J and  Rogers G 1982 {\it Phys. Rev. Lett.} \href{https://doi.org/10.1103/PhysRevLett.49.1804}{ {\bf 49} 1804} }

\bibitem{zelliger}{ Artur B D E and Anton Z 2000  \textit{The Physics of Quantum Information: Quantum Cryptography, Quantum Teleportation, Quantum Computation} (Oxford: Springer Publishing Company) }
  

\bibitem{yurker1986}{ Yurke B and  Stoler D 1986 \textit{Phys. Rev. Lett.} \href{https://journals.aps.org/prl/abstract/10.1103/PhysRevLett.57.13}{ {\bf 57} 13   }}
  

\bibitem{yurker1987}{ Yurke B and  Stoler D 1987  \textit{Phys. Rev. \emph{A}} \href{https://journals.aps.org/pra/abstract/10.1103/PhysRevA.35.4846}{ {\bf 35}  4846 }}


\bibitem{Sanders}{ Sanders B C 1992 {\it Phys. Rev. \emph{A}} \href{https://journals.aps.org/pra/abstract/10.1103/PhysRevA.45.6811}{ {\bf 45} 6811}}


\bibitem{leibfried}{ Leibfried D et. al.  2005 \textit{Nature} \href{https://www.nature.com/articles/nature04251}{ \textbf{438}  639}}


\bibitem{hirota2001}{Hirota O and  Sasaki M 2002 \textit{Quantum Communication, Computing, and Measurement 3} (Boston: Springer), p. 359-366.}

\bibitem{gilchrist12}{ Gilchris A,  Nemoto K,  Munro W J,  Ralph T C,  Glancy S,  Braunstein S L and  Milburn G J 2004  \textit{J. Opt. B: Quantum Semiclass. Opt.} \href{https://iopscience.iop.org/article/10.1088/1464-4266/6/8/032/pdf}{ \textbf{6} S828 }}


\bibitem{Hirota2001Decoherence}{ Hirota O,  Enk S J,  Nakamura K, Sohma M and  Kato K 2001 Entangled Nonorthogonal States and Their Decoherence Properties \href{https://arxiv.org/abs/quant-ph/0101096}{ { (arXiv:quant-ph/0101096)}}}


\bibitem{Hirota2001Teleportation}{ Enk S J and  Hirota O 2001 {\it Phys. Rev. \emph{A}} \href{https://journals.aps.org/pra/abstract/10.1103/PhysRevA.64.022313}{{\bf 64} 022313}}


\bibitem{Kim2001}{ Jeong H,  Kim M S and  Lee J 2001 {\it Phys. Rev. \emph{A}} \href{https://journals.aps.org/pra/abstract/10.1103/PhysRevA.64.052308}{ {\bf 64} 052308 }}


\bibitem{Wang}{ Wang X 2001 {\it Phys. Rev. \emph{A}} \href{https://journals.aps.org/pra/abstract/10.1103/PhysRevA.64.022302}{ {\bf 64} 022302 }}


\bibitem{Kim2002}{ Jeong H and  Kim M S 2002 \textit{Phys. Rev. \emph{A}} \href{https://journals.aps.org/pra/abstract/10.1103/PhysRevA.65.042305}{ \textbf{65} 042305 } }  


\bibitem{Ralph}{ Ralph T C,  Munro W J, and  Milburn G J 2002 \textit{Quantum Optics in Computing and Communications} \href{https://doi.org/10.1117/12.483016}{ \textbf{4917} 1}}


\bibitem{HirotaQuantumFreeError}{ Hirota O 2011 Error free Quantum Reading by Quasi Bell State of Entangled Coherent States \href{https://arxiv.org/abs/1108.4163}{(arXiv:1108.4163)}}


\bibitem{Hirota2016}{ Hirota O 2016 {\it J. Laser Opt. Photonics} \href{https://doi.org/10.4172/2469-410X.1000129}{ {\bf 3} 129  }}


\bibitem{Kato}{ Kato K 2015 {\it Quantum Communications and Quantum Imaging XIII} \href{https://doi.org/10.1117/12.2188103}{ \textbf{9615} 96150N }}

\bibitem{UmezawaLivro}{Umezawa H 1995 {\it Advanced Field Theory: Micro, Macro, and Thermal} Physics (New York: AIP Press)}


\bibitem{takahashi}{ Takahashi Y and  Umezawa H 1996 \textit{Int. J. Mod. Phys. \emph{B}} \href{https://www.worldscientific.com/doi/abs/10.1142/S0217979296000817}{ \textbf{10} 1755 }}

\bibitem{Umezawa}{ Takahashi Y and  Umezawa H 1975 {\it Collec. Phen.} {\bf 2} 55}


\bibitem{trindade}{ Trindade M A S, Silva Filho L M , Santos L C, Martins M G R and Vianna J D M 2013 \textit{Int. J. Mod. Phys. \emph{B}} \href{https://doi.org/10.1142/S0217979213501336}{ \textbf{27} 1350133 }}


\bibitem{floquet}{ Floquet S, Trindade  M A S and  Vianna J D M 2017 \textit{Int. J. Mod. Phys. \emph{A}} \href{https://doi.org/10.1142/S0217751X17500154}{ \textbf{32}  1750015 }}


\bibitem{kitajima}{ Kitajima S, Arimitsu T, Obinata M and Yoshida K 2014  \textit{Physica \emph{A}} \href{https://doi.org/10.1016/j.physa.2014.02.068}{ \textbf{404} 242}}

\bibitem{prudencio}{ Prud\^encio T,  Rocha Filho T M and  Santana A E 2019 \textit{Physica Scripta} \href{https://iopscience.iop.org/article/10.1088/1402-4896/ab03b1/meta}{  \textbf{94} 095102}}


\bibitem{prudencio2}{ Prud\^encio T 2012 \textit{Int. J. Quant. Inf.} \href{https://doi.org/10.1142/S021974991230001X}{ \textbf{10} 1230001}}


\bibitem{zhang}{ Zhang W, Feng D H and Gilmore R 1990 \textit{Rev. Mod. Phys.} \href{https://journals.aps.org/rmp/abstract/10.1103/RevModPhys.62.867}{ \textbf{62} 867}}

\bibitem{Wigner}{Wigner E P 1932 {\it Phys. Rev. } \href{https://journals.aps.org/pr/abstract/10.1103/PhysRev.40.749}{ {\bf 40} 749 }}



\bibitem{bell87}{ Bell J S 1987 \textit{Speakable and Unspeakable in Quantum Mechanics} (Cambridge: Cambridge University Press) pp 196--200.}

\bibitem{mc15}{ McConnell R,  Zhang H,  Hu J,  \'Cuk S and  Vuleti\'c V 2015 \textit{Nature} \href{https://www.nature.com/articles/nature14293}{ {\bf 519} 439 }}

\bibitem{smi93}{ Smithey D T, Beck  M,  Raymer M G, and  Faridani A 1993 \textit{Phys. Rev. Lett.} \href{https://journals.aps.org/prl/abstract/10.1103/PhysRevLett.70.1244}{ {\bf 70}, 1244}}

\bibitem{ku97}{ Kurtsiefer C,  Pfau T and  Mlynek J 1997 \textit{Nature} \href{https://www.nature.com/articles/386150a0}{ {\bf 386} 150 }}


\bibitem{integrais}{ Gradshteyn I S,  Ryzhik I M and Romer  R H 1988 \textit{Tables of integrals, series, and products}  (San Diego: Elsevier)}

\bibitem{butkov}{ Butkov E 1973 \textit{Mathematical Physics} (London: Addison-Wesley Publishing)}


\bibitem{wang2016}{ Wang C 2016 \textit{Science} \href{https://science.sciencemag.org/content/352/6289/1087}{ {\bf 352} 1087 .}}


\bibitem{vla2015}{ Vlastakis B et al. 2015 \textit{ Nat. Commun.} \href{https://www.nature.com/articles/ncomms9970/}{ {\bf 6} 1}}


\bibitem{song2019}{ Song C  et al. 2019 \textit{Science} \href{https://science.sciencemag.org/content/365/6453/574.abstract}{ {\bf 365} 574 }}

\bibitem{rcran}{R Core Team 2013 \textit{R: A Language and Environment for Statistical Computing} (Vienna: R Foundation for Statistical Computing)}


\bibitem{weinbub2018}{ Weinbub J and  Ferry D K 2018 \textit{ Appl. Phys. Rev.} \href{https://aip.scitation.org/doi/full/10.1063/1.5046663}{  {\bf 5} 041104 }}



\bibitem{kenfack2004}{ Kenfack A,  \.{Z}yczkowski K 2004 \textit{J. Opt. B: Quantum  Semiclass. Opt.} \href{https://iopscience.iop.org/article/10.1088/1464-4266/6/10/003/meta}{ {\bf 6} 396}}


\bibitem{alexei}{ Ourjoumtsev A,   Jeong H,  Tualle-Brouri R and  Grangier P 2007 \textit{Nature} \href{https://www.nature.com/articles/nature06054}{ \textbf{448} 784}}

\bibitem{benedict99}{ Benedict M G and  Czirjak A 1999 \textit{Phys. Rev. \emph{A}} \href{https://journals.aps.org/pra/abstract/10.1103/PhysRevA.60.4034}{ {\bf 60} 4034}}

\bibitem{foldi2002}{ F\"oldi P,  Czirják A,  Moln\'ar B, and  Benedict M G 2002 \textit{ Opt. Express} \href{https://doi.org/10.1364/OE.10.000376}{ {\bf 10} 376}}
\end{thebibliography}
\end{document}